\documentclass[journal]{IEEEtran}

\usepackage{amssymb}
\usepackage{amsfonts}
\usepackage{graphicx}
\usepackage[cmex10]{amsmath}
\usepackage{times}
\usepackage{booktabs}
\usepackage{subfigure}
\usepackage{comment}
\usepackage{color}

\usepackage{algorithm} 
\usepackage{algorithmic} 
\usepackage{amsmath}

\newcommand{\D}{\mathcal{D}}

\renewcommand{\S}{s}
\newcommand{\DC}{\textrm{DC}}
\newcommand{\Beta}{\pmb{\beta}}

\newtheorem{lemma}{Lemma}
\newtheorem{theorem}{Theorem}
\newtheorem{definition}{Definition}

\begin{document}


\title{Accelerating  Data Regeneration for Distributed Storage Systems with Heterogeneous Link Capacities}

\author{Yan~Wang,
\thanks{Y. Wang is with the School of Software, East China Jiao Tong University, Nanchang, China. The main work was done when she was with the School of Computer Science, Fudan
University, email:ywangc11@fudan.edu.cn}
		Xunrui Yin,
\thanks{X. Yin is with Department of Computer Science, University of Calgary, Canada.}
		Dongsheng Wei,
		Xin Wang, ~\IEEEmembership{Member,~IEEE,}
\thanks{D. Wei and X. Wang are with the School of Computer Science, Fudan
University.}
		Yucheng He, ~\IEEEmembership{Member,~IEEE}
\thanks{Y. He is with the School of Information Science and Engineering, Huaqiao University, Xiamen, China.}
\thanks{This paper was presented in part at IEEE INFOCOM 2014.}
}

\maketitle

\begin{abstract}
Distributed storage systems provide large-scale reliable data storage services by spreading redundancy across a large group of storage nodes. In such a large system, node failures take place on a regular basis. When a storage node breaks down, a replacement node is expected to regenerate the redundant data as soon as possible in order to maintain the same level of redundancy.
Previous results have been mainly focused on the minimization of network traffic in regeneration. However, in practical networks, where link capacities vary in a wide range, minimizing network traffic does not always yield the minimum regeneration time.
In this paper, we investigate two approaches to the problem of minimizing regeneration time in networks with heterogeneous link capacities. The first approach is to download different amounts of repair data from the helping nodes according to the link capacities. The second approach generalizes the conventional star-structured regeneration topology to tree-structured topologies so that we can utilize the links between helping nodes with bypassing low-capacity links.
Simulation results show that the flexible tree-structured regeneration scheme that combines the advantages of both approaches can achieve a substantial reduction in the regeneration time.
\end{abstract}


\section{Introduction}
\label{sec: introduction}

Large-scale distributed storage systems are widely used today to provide reliable data storage services, by spreading data redundancy over a large number of storage nodes. Users can access their data anytime and anywhere. Such application scenarios include large data centers such as Google File Systems, Total Recall, OceanStore and peer-to-peer storage systems Wuala. 

In distributed storage systems, the level of redundancy is usually described by parameters $(n,k)$, where $n$ is the number of storage nodes holding coded blocks of a file, and $k$ indicates that the file can be reconstructed from any $k$ out of the $n$ storage nodes. For example, we can use an $(n,k)$ maximum distance separable (MDS) code to encode a file of size $M$ into $n$ blocks of equal size $M/k$, and disseminate them to $n$ storage nodes with each holding one block. Then, the original file can be reconstructed from any  $k$ storage nodes. In literature, the ability to reconstruct the file from any $k$ storage nodes is usually referred to as the {\em MDS property} \cite{nccloud}.

In such big systems, nodes fail frequently, and the failures should be handled on a routine basis. After a node fails or leaves the system, the reliability degrades, and the protected data becomes vulnerable. To maintain the same level of redundancy, it is important to regenerate the lost data at a replacement node, called {\em newcomer}, as soon as possible \cite{gfs}. In this work, we focus our attention on how to minimize regeneration time in the regeneration process.

An intuitive way to achieve a minimized regeneration time is to minimize the total amount of data transferred for regeneration, which is called {\em repair bandwidth}. In this direction, Dimakis {\em et al.} proposed Regenerating Codes to achieve the minimum repair bandwidth \cite{Dimakis2010}. For simplicity, they assumed that each surviving node participating in the regeneration, called {\em provider}, sends the same amount of repair data to the newcomer. In this paper, we focus on the regenerating codes with functional repair, where the regenerated data may be different from the lost data. We will use the term {\em repair traffic} to denote the amount of repair data transmitted from each provider to the newcomer. 

Under the assumption of uniform repair traffic, Demakis {\em et al.} derived the minimum repair bandwidth to maintain the MDS property.
They found that there is a trade-off between repair bandwidth and storage efficiency, with two extremal cases: 1) the {\em minimum-storage regenerating} (MSR) point where each node stores the minimum amount of data, and 2) the {\em minimum-bandwidth regenerating} (MBR) point where the storage efficiency is sacrificed for achieving the minimum repair bandwidth.

In practice, minimum repair bandwidth does not always result in minimum regeneration time, especially in heterogeneous networks where link capacities vary in a wide range. The heterogeneous networks are commonly deployed for distributed storage systems. For example, within a  data center, servers are usually placed in racks, and the servers in the same rack may enjoy a much larger bandwidth than those located in different racks \cite{realisticRackModel}. Meanwhile, the available bandwidths among servers are distinct because of different background traffic, even if the link capacities are the same \cite{trafficCharacteristic}. The difference of link bandwidths becomes even larger when using multiple geo-distributed data centers to safeguard
users' data from the failure of an entire data center, which is a
conventional practice for large companies, {\em e.g.} Google.
In these networks, the regeneration time depends not only on the repair bandwidth, but also on the bandwidths of bottleneck links between providers and the newcomer.
Thus, in this work, we will also consider nonuniform repair traffic in order for the amount of transmitted repair data to match the available bandwidth over heterogeneous links.

There are, in general, two approaches for accelerating the regeneration process in heterogeneous networks. The first is to drop the assumption of uniform repair traffic. Instead, the repair traffic would better be dynamically determined according to the end-to-end available bandwidth from each provider to the newcomer. The idea is to let providers with larger end-to-end available bandwidths transmit more repair data for reducing in turn the amount of data transmitted through bottleneck links. This approach requires a new form of restriction on the repair traffic to maintain the MDS property. For example, Shah {\em et al.} \cite{FlexibleRC_ISIT2010} developed a regenerating scheme that supports non-uniform repair traffic with two newly introduced parameters $\beta_{max}$ and $\gamma$. While the repair traffic can be dynamically determined for each provider in each regeneration process, it requires that the provider transmits at most an amount of $\beta_{max}$ repair data and the total amount of repair data from all the providers must be no less than $\gamma$, so that the MDS property can be preserved.

The second approach is to utilize the inter-provider links to bypass bottleneck links between providers and the newcomer. This idea was first proposed by Li {\em et al.} \cite{tree}. They designed a {\em tree-structured} regeneration scheme (called {\em RCTREE}), where the tree has the newcomer as the root and has providers as intermediate and leaf nodes, and the repair data reach the newcomer along the branches of the tree. In addition, they allowed the repair data to be encoded at the intermediate nodes of the tree to further reduce the regeneration time. However, because of insufficient amount of repair data transmitted, RCTREE cannot maintain the MDS property.


In this paper, we study the problem of minimizing regeneration time in heterogeneous networks while maintaining the MDS property. Our contributions include two parts as described below.

First, we propose a regenerating scheme supporting non-uniform repair traffic. Compared with previous studies, our scheme has the advantage of choosing the repair traffic according to the set of largest available bandwidths while maintaining the MDS property, and thus achieves the best possible regeneration time among the schemes those are based on dynamic determination of repair traffic. We introduce the {\em feasible region} to generalize the restrictions on the repair traffic for the MDS property. For the MSR case, we study the structure and the uniqueness of maximal feasible regions and then find their optimal solution. For the non-MSR case, we construct a heuristic feasible region with which the repair traffic can be dynamically determined by solving a linear programming problem for each round of repair. We refer to this regeneration scheme as flexible regeneration (FR).

Second, we reconsider the tree-structured regeneration approach with regenerating codes, because the previous tree-structured regeneration scheme \cite{tree} cannot preserve the MDS property in the repair process. We develop the information flow graph method from \cite{Dimakis2010}.
Using this method, we derive the minimum amount of data to be transmitted on each link in a given regeneration tree, and formulate the problem of building an optimal regeneration tree to minimize the regeneration time. Unfortunately, we find this problem NP-complete, mainly because the information flow on each link exhibits a correlation with the number of providers using this link. We thus propose a heuristic algorithm, called Tree-structured Regeneration (TR), to find a near-optimal regeneration tree. Furthermore, we propose a Flexible Tree-structured Regeneration (FTR) scheme by combining TR with FR.

\begin{figure*}
\centering
\includegraphics[width=0.85\textwidth]{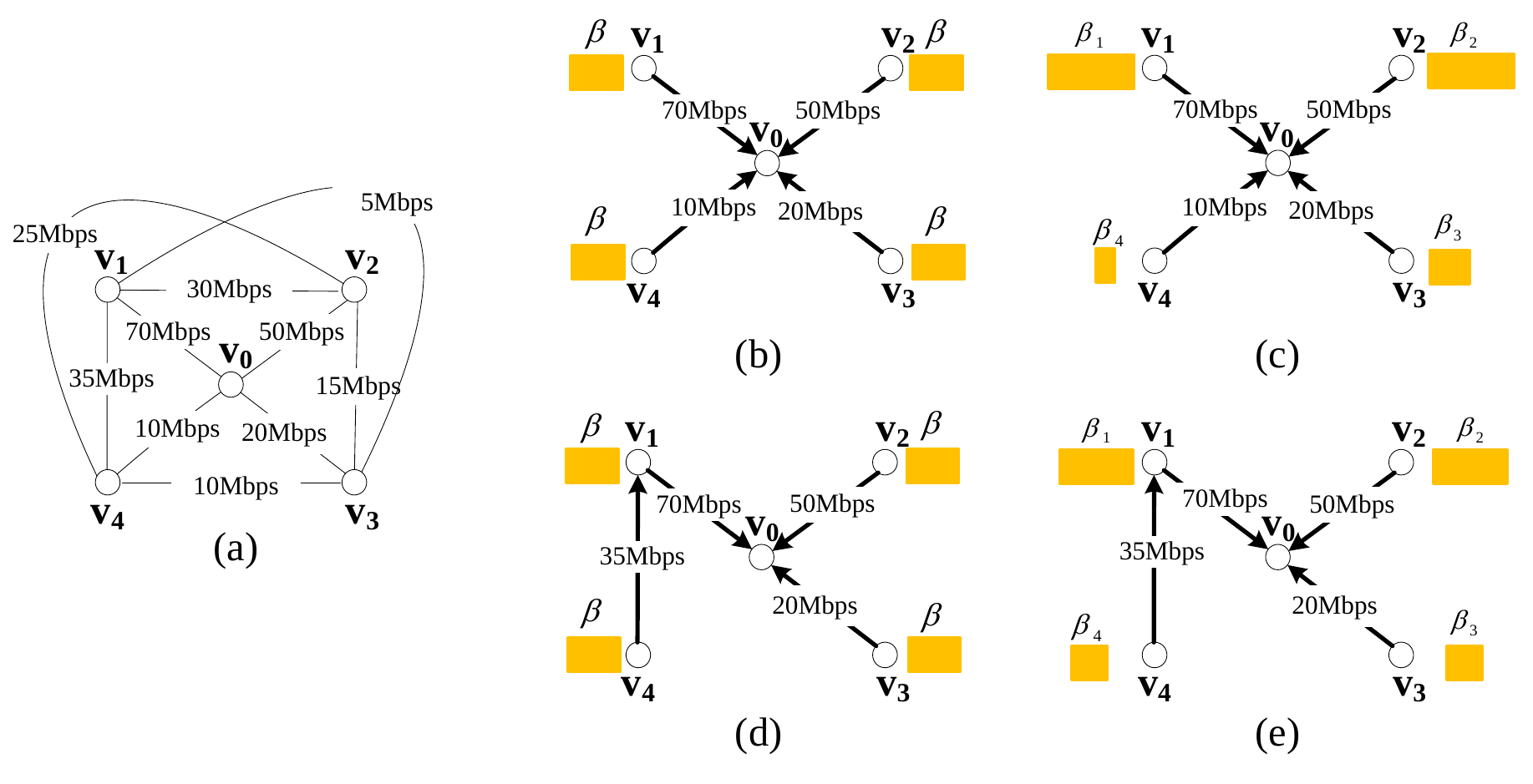}
\caption{Examples for tree regeneration schemes: STAR, FR, TR and FTR. The parameters are $n=5$, $d=4$, $k=2$, $M=480$Mb, $\alpha=M/k=240$Mb, $\beta=\frac{\alpha}{d-k+1}=80$Mb. The lengths of the blocks indicate the amount of repair data generated from each provider. The regeneration time of STAR, FR, TR and FTR is 8s, 4s, 3s and 2.67s, respectively.}\label{example}
\end{figure*}

These ideas can be illustrated with the example shown in Fig.~\ref{example}. Consider the overlay network shown in Fig.~\ref{example}(a), where  $v_0$ is the newcomer, and $ v_1, v_2, v_3, v_4$ are the $d=4$ providers. The bandwidths are labeled on the links, which range from $5$Mbps to $70$Mbps. We assume that the redundancy is coded as an $(n=5,k=2)$-MDS code, such that any $2$ out of $5$ storage nodes are able to reconstruct the file. Suppose that the size of the original file is equal to $M=480$Mb, and each storage node stores $\alpha=M/k=240$Mb.

Fig.~\ref{example}(b) shows the conventional regeneration scheme (STAR)\cite{Dimakis2010} which uses the star-structured topology: $v_0$ receives data directly from the four providers. In order to regenerate the lost data at $v_0$ using the regenerating code, an amount of $\beta=\frac{M}{k(d-k+1)}=80$Mb data need to be downloaded from each provider.
Thus, the regeneration time of STAR is determined by the slowest transmission which takes $\frac{\beta}{10Mbps}=8$ seconds.
Fig.~\ref{example}(c) shows the STAR-based flexible regeneration (FR) scheme, where each provider is allowed to generate a different amount of coded data. We find that the MDS property can be maintained if providers $v_1, v_2, v_3, v_4$ generate  and send $\beta_1 = 150$Mb, $\beta_2 = 150$Mb, $\beta_3 = 60$Mb, $\beta_4 = 30$Mb, respectively.
As a result, the regeneration time of FR is reduced to $\max\{\frac{\beta_1}{70Mbps}, \frac{\beta_2}{50Mbps}, \frac{\beta_3}{20Mbps}, \frac{\beta_4}{10Mbps}\} = 3$ seconds.

The regeneration trees used by TR and FTR are shown in Fig.~\ref{example}(d) and (e), respectively. They coincide with the same tree in this example. In TR, each provider generates $\beta=80Mb$ coded data with its local storage. According to our analysis, the minimum amount of data transferred on edge $(v_1,v_0)$ is at least $2\beta$. Thus, it costs TR $\max\{\frac{2\beta}{70Mbps}, \frac{\beta}{50Mbps}, \frac{\beta}{20Mbps}, \frac{\beta}{35Mbps}\}=4$ seconds to accomplish the regeneration. With FTR, each provider may generate different amount of repair data. The solution given by our FTR algorithm is that providers $v_1, v_2, v_3, v_4$ generate $\beta_1 = 133.33$Mb, $\beta_2 = 133.33$Mb, $\beta_3 = 53.33$Mb, $\beta_4 = 53.33$Mb,  respectively. The amount of data flow transmitted on the link $(v_1, v_0)$ is $\beta_1 + \beta_4 = 186.67$Mb, and the regeneration time is $\max\{\frac{\beta_1+\beta_4}{70Mbps}, \frac{\beta_2}{50Mbps}, \frac{\beta_3}{20Mbps}, \frac{\beta_4}{35Mbps}\} = 2.67$ seconds.

To evaluate the performances of our schemes, we implement these regeneration schemes and carry out simulations with practical link capacities. Simulation results show that, depending on heterogeneous link capacities, our proposed schemes can reduce the regeneration time by $10\%\sim 90\%$, compared with the conventional STAR-structured regeneration scheme.


The remainder of the paper is organized as follows.
In Section~\ref{sec: formulation}, we formulate the process of data regeneration in distributed storage systems and introduce the random linear coding and information flow graph, which is the theoretical tool applied in the analysis.
In Section~\ref{sec:fr}, we study the flexible regeneration.
In Section~\ref{sec:tree}, we reconsider the tree-structured regeneration scheme by analyzing the corresponding information flow graph.
In Section~\ref{sec: flexible}, we propose a flexible scheme based on tree-structured regeneration.
In Section~\ref{sec: evaluation}, we evaluate the performance of our three proposed regeneration schemes. Finally, we introduce the related works in Section~\ref{sec: relate}, and conclude the paper by Section~\ref{sec: conclusion}.

\section{Problem Formulation}
\label{sec: formulation}
From this section, we expand the unit from 'bit' to 'block' that comprises a specific length of bits for describing the variables $M, \alpha, \beta$, and $\beta_i$.
 
Assume that a file is divided into $M$ blocks and encoded into $n\alpha$ blocks of equal length. The coded blocks are disseminated to $n$ storage nodes, with each node holding $\alpha$ blocks. The reliability requirement is formalized as the MDS property, which requires that the file can be reconstructed by accessing any $k$ storage nodes. After a storage node fails, the newcomer accesses $d$ providers to regenerate $\alpha$ blocks. With regenerating codes, each provider generates $\beta$ coded blocks by encoding the local $\alpha$ blocks and directly transmits them to the newcomer.

We use a complete graph $G(V,E)$ to represent the overlay network consisting of the $d$ providers and the newcomer \cite{tree}. The $d$ provider-to-newcomer flows form a star topology centered at the newcomer. For any two nodes $u,v\in V$, let $c(u,v)$ denote the link capacity from $u$ to $v$. For a specific regeneration process, let $f(u,v)$ denote the number of blocks transmitted over the link $(u,v)$. We assume that the coding operations are streamlined with the data transmission, which dominates the regeneration time. Notice that in real-world overlay networks, the end-to-end links usually have different capacities, which may vary in one or two orders of magnitude \cite{measuring}. 
The regeneration time can be simply represented  as
\[\max\left\{\frac{f(u,v)}{c(u,v)} | (u,v)\in E\right\}\]

The challenge is how to minimize the regeneration time without violating the MDS property. The regeneration process based on the star topology may suffer from a bottleneck caused by the lowest link capacity between the providers and the newcomer. In general, we can utilize two approaches to accelerate the regeneration process in heterogeneous networks. The first is to generate a different amount of repair data from each provider according to its outgoing link capacity. The second is to utilize the links between providers to by-pass links of low capacities. 

Therefore, a regeneration scheme can be decomposed into two parts: 1)  an algorithm that flexibly determines the repair flows $f(u,v), (u,v)\in E$, while assuring that the newcomer obtains enough information to regenerate the desired data; 2) a set of codes indicating how to encode the blocks and how to reconstruct the file. For the latter part, we employ the random linear network codes, with which we are able to utilize the information flow graph technique \cite{Dimakis2010} to simplify the problem.

\subsection{Random linear network coding and information flow graph}

Dimakis {\em et al.} \cite{Dimakis2010} proposed the information flow graph technique to analyze the minimum repair bandwidth for maintaining the MDS property. Specifically, they construct the information flow graph in the following way. For each storage node $u$, create two nodes $u^{in}$ (in-node) and $u^{out}$ (out-node) and a link of capacity $\alpha$ from $u^{in}$ to $u^{out}$. Create a source node $s$ and for each initial storage node $u$, add a link from $s$ to $u^{in}$ with  infinite capacity. For a storage node $v$ regenerated by accessing $d$ providers, add $d$ links from the out-nodes of the $d$ providers to the in-node of newcomer $v$, each with capacity $\beta$. For the purpose of proposal in this paper, we assume various link capacities $\beta_i$ between the providers and the newcomer. If a data collector(DC) connects to $k$ storage nodes to reconstruct the file, add $k$ links of infinite capacity from the out-nodes of these storage nodes to the data collector.

%

In the distribution process, the original file is divided into $M$ blocks $a_j (j=1,2, \cdots, M)$ of equal length, and is encoded into $n\alpha$ blocks $b_i (i=1,2,\cdots,n\alpha)$. These coded blocks are then evenly distributed to $n$ storage nodes. With linear codes, each coded block $b_i$ is a linear combination of blocks $a_1, a_2, \cdots, a_M$, {\em i.e.,}
\[
\left[\begin{array}{c}
b_1\\
b_2\\
\vdots\\
b_{n\alpha}
\end{array}
\right]
 =
 \left[\begin{array}{cccc}
c_{1,1} & c_{1,2} & \cdots & c_{1,M}\\
c_{2,1} & c_{2,2} & \cdots & c_{2,M}\\
\vdots  & \vdots  & \ddots & \vdots\\
c_{n\alpha,1} & c_{n\alpha,2} & \cdots & c_{n\alpha, M}\\\end{array}
\right]
\left[\begin{array}{c}
a_1\\
a_2\\
\vdots\\
a_M
\end{array}
\right]
\]
For the MDS property, we have to carefully choose the combination coefficients $c_{i,j}$ so that the original blocks $a_j$ can be reconstructed from any $k$ storage nodes. A possible encoding method is to utilize Reed-Solomon codes \cite{Reed1960}, {\em e.g.}, setting the generator matrix $[c_{i,j}]$ to be a Vandermonde matrix. For ease of implementation, the combination coefficients $c_{i,1}, c_{i,2}, \cdots, c_{i,M}$ for any specific coded block $b_i=\sum_{j=1}^M c_{i,j} a_j$ are transmitted and stored together with block $b_i$. For simplicity, we use $\vec{c} = (c_{1}, c_{2}, \cdots, c_{M})$ to denote the row vector of combination coefficients, which is also called the coding vector of block $b$.

In the regeneration process, we employ a random linear network coding scheme as the regenerating code to facilitate the regeneration of coded blocks at the newcomer. Without loss of generality, each provider $v_i$ generates and transmits $\beta_i$ coded blocks $b^{(r)}_{i,1},b^{(r)}_{i,2},\cdots,b^{(r)}_{i,\beta_i}$ as random
linear combinations of its local $\alpha$ blocks
$b_{i,1},b_{i,2},\cdots,b_{i,\alpha}$ as follows
\[
\left[\begin{array}{c}
b^{(r)}_{i,1}\\
b^{(r)}_{i,2}\\
\vdots\\
b^{(r)}_{i,\beta_i}
\end{array}
\right]
=
\left[\begin{array}{cccc}
c^{(r)}_{1,1}&c^{(r)}_{1,2}&\cdots&c^{(r)}_{1,\alpha}\\
c^{(r)}_{2,1}&c^{(r)}_{2,2}&\cdots&c^{(r)}_{2,\alpha}\\
\vdots&\vdots&\ddots&\vdots\\
c^{(r)}_{\beta_i,1}&c^{(r)}_{\beta_i,2}&\cdots&c^{(r)}_{\beta_i,\alpha}\\
\end{array}
\right] \left[\begin{array}{c}
b_{i,1}\\
b_{i,2}\\
\vdots\\
b_{i,\alpha}
\end{array}
\right]
\] where $i=1,2,\cdots,d$, and the coefficients $c^{(r)}_{i,j}$ are randomly chosen from the finite field of the linear regenerating code such that the coefficient matrix $[c^{(r)}_{i,j}]$ has rank $\beta_i$.

To complete the regeneration process, the newcomer receives $\gamma = \sum_{i=1}^d \beta_i$ blocks and generates its own $\alpha$ blocks by
\[
\left[\begin{array}{c}
b'_1\\
b'_2\\
\vdots\\
b'_{\alpha}
\end{array}
\right]
 =
 \left[\begin{array}{cccc}
c'_{1,1} & c'_{1,2} & \cdots & c'_{1,\gamma}\\
c'_{2,1} & c'_{2,2} & \cdots & c'_{2,\gamma}\\
\vdots  & \vdots  & \ddots & \vdots\\
c'_{\alpha,1} & c'_{\alpha,2} & \cdots & c'_{\alpha, \gamma}\\\end{array}
\right]
\left[\begin{array}{c}
b^{(r)}_{1,1}\\
\vdots\\
b^{(r)}_{1,\beta_1}\\
b^{(r)}_{2,1}\\
\vdots\\
b^{(r)}_{2,\beta_2}\\
\vdots\\
b^{(r)}_{d,\beta_d}
\end{array}
\right]
\]
where the coefficient matrix $[c'_{i,j}]$ is
obtained from the coefficient matrices $[c^{(r)}_{i,j}]$ for the linear regenerating code. It should be noted that the coding vectors of the blocks $b'_i$ are also similarly generated and stored at the newcomer.

In the reconstruction process, the data collector collects $k\alpha$ blocks $b''_1, b''_2, \cdots, b''_{k\alpha}$ from $k$ storage nodes, whose coding vectors form a $k\alpha$-by-$M$ matrix $[c''_{i,j}]$. As long as the matrix has rank $M$, we can reconstruct the original file from these coded blocks (as required by the MDS property \cite{nccloud}) by solving the following linear equation:
\[
\left[\begin{array}{c}
b''_1\\
b''_2\\
\vdots\\
b''_{k\alpha}
\end{array}
\right]
 =
 \left[\begin{array}{cccc}
c''_{1,1} & c''_{1,2} & \cdots & c''_{1,M}\\
c''_{2,1} & c''_{2,2} & \cdots & c''_{2,M}\\
\vdots  & \vdots  & \ddots & \vdots\\
c''_{k\alpha,1} & c''_{k\alpha,2} & \cdots & c''_{k\alpha, M}\\\end{array}
\right]
\left[\begin{array}{c}
a_1\\
a_2\\
\vdots\\
a_M
\end{array}
\right]
\]

With the information flow graph, Dimakis {\em et al.} proved the following result \cite{Dimakis2010}:
\begin{lemma}\label{lem1}
In the information flow graph constructed according to the repair history of a distributed storage system, if the max-flow from $\S$ to a data collector is no less than the file size $M$, then with random linear codes over field $\mathbb{F}$, the data collector can recover the file with probability arbitrarily close to 1 as $|\mathbb{F}| \rightarrow \infty$.
\end{lemma}



\section{Star-structured Regeneration with Flexible Repair Traffic}
\label{sec:fr}
We first consider the approach of utilizing flexible non-uniform repair traffic to accelerate the regeneration process with heterogeneous link capacities. Generalizing the set of repair traffic that preserves the MDS property as the ``feasible region'', we characterize the structure of maximal feasible regions and derive the flexible regeneration scheme whose feasible region subsumes the feasible region of previous studies.

Specifically, for a regeneration process, let $\beta_i$ denote the repair traffic, {\em i.e.}, the number of blocks transmitted from the $i$-th provider to the newcomer, and let $\Beta = (\beta_1, \beta_2, \cdots, \beta_d)$ denote the repair bandwidth in terms of the vector of repair traffic. For each round of repair, if we have multiple choices of $\Beta$, then we say that `flexible repair traffic is supported'. Let $\D$ denote the set of possible choices of $\Beta$. We call $\D\subset \mathcal{R}^d$ a feasible region, if the MDS property is maintained as long as $\Beta$ is chosen from $\D$.

With a given feasible region, minimizing the regeneration time in each round of repair is equivalent to solving:
\begin{eqnarray}
\min_{\Beta\in \D} \max_{i=1,\cdots,d} \frac{\beta_i}{c(v_i,v_0)}
\label{optimization1}
\end{eqnarray}
where $v_i$ is the $i$-th provider and $v_0$ is the newcomer. The link capacities $c(v_i, v_0)$ may be simply written as $c_i$.

According to the analysis of feasible regions in the following subsection, we can see that a maximal feasible region is actually a convex polytope, which makes problem (\ref{optimization1}) a linear programming problem and solvable in polynomial time.

\subsection{Structure of a maximal feasible region}
In the information flow graph, the capacities of links entering a newcomer's in-node represent the amount of information downloaded from each provider. Therefore, with flexible repair traffic, we set the corresponding link capacities as $\beta_i, i=1,\cdots,d,$ instead of $\beta$. With this minor difference from the conventional information flow graph, it can be seen that Lemma \ref{lem1} still holds.
\label{sec: structure}
\begin{figure*}[t]
	\centering
	 \includegraphics[width=.7\textwidth]{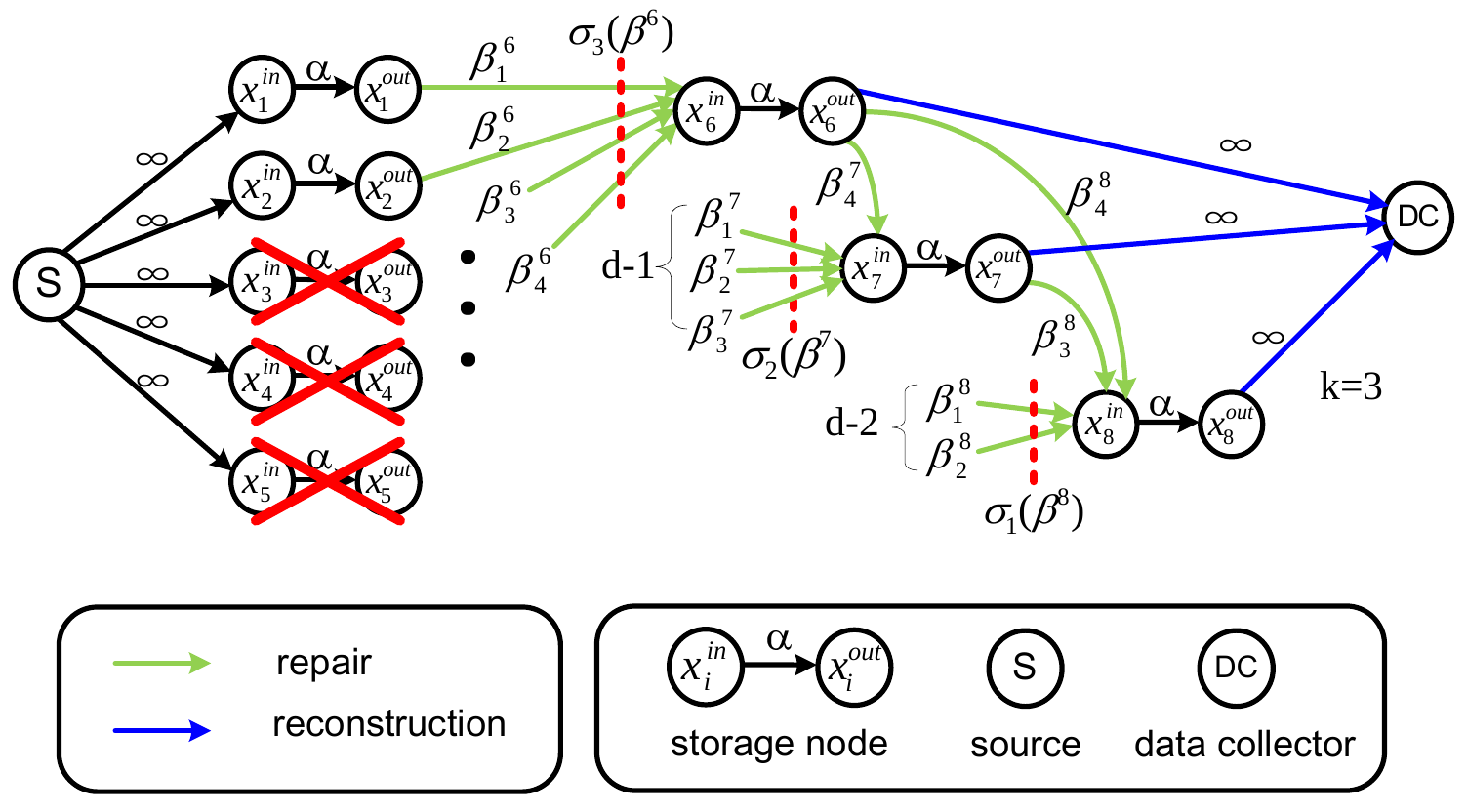}
	\caption{An example of the information flow graph when $k=3, d=4$, where $\Beta^i$ denotes the repair bandwidth in the repair when the node $x^i_{in}$ acts as the newcomer. Without loss of generality, we suppose that the elements of $\Beta^i$ are sorted such that $\beta^i_1 \leq \beta^i_2 \leq \beta^i_3 \leq \beta^i_4, i=0,1,2$, and let $\sigma_j(\Beta^i) = \sum_{l=1}^{d-k+j}\beta^i_l$. We can see that the min-cut equals $\sum_{j=1}^k \min\{\sigma_j(\Beta^{9-j}),\alpha\}$ in this figure.}
	\label{fig: flowgraph}
\end{figure*}

In order to convert the MDS property into constraints on the feasible region, we analyze the min-cuts of the information flow graph:
\begin{lemma}
If in each round of repair, the newcomer accesses $d$ providers with the repair bandwidth $\Beta \in \D$. Then in the corresponding information flow graph, the min-cut between $\S$ and any data collector DC satisfies
\begin{eqnarray}
\textrm{min-cut}(\S,\DC) \geq \sum_{j=1}^{k} \min \left\{ \min_{\Beta \in \mathcal{D}}\sigma_j(\Beta),\alpha \right\}
\label{eq: model: min-cut}
\end{eqnarray}
Here $\sigma_j(\Beta)$ is defined as the sum of the $d-k+j$ smallest numbers of $\beta_1,\beta_2,\dots,\beta_d$. This bound can be matched with equality.
\end{lemma}
\begin{IEEEproof}
Let $(U,\bar{U})$ denote a cut separating $\DC$ and $\S$ such that $\DC\in U$ and $\S \in \bar{U}$, respectively. Then $U$ must contain at least $k$ storage nodes. Consider each of the first $k$ nodes of $U$ in the topological sorting. If it is an input node, the set of repair links will be included in the cut. If it is an output node, the storage link of capacity $\alpha$ will be included. As illustrated in Fig.~\ref{fig: flowgraph}, it can be verified that the equality of (\ref{eq: model: min-cut}) can be achieved when there are $k$ cascading repairs, each of which includes the newcomers in previous rounds as helper nodes.
\end{IEEEproof}

As the bound of the min-cut (\ref{eq: model: min-cut}) can be matched with equality, we use $MC(\mathcal{D},\alpha)$ to represent the minimum min-cut of all possible information flow graphs with a given region $\D$ and the storage per node $\alpha$. If we want to ensure that any $k$ storage nodes suffice to reconstruct the original file, while allowing a newcomer to connect to any set of $d$ providers, then the following condition must be satisfied:
\begin{equation}\label{mc}
   MC(\D,\alpha) = \sum_{j=1}^{k} \min \left\{ \min_{\pmb{\beta} \in \mathcal{D}}\sigma_j(\pmb{\beta}),\alpha \right\}\geq M.
\end{equation}
We refer to this condition as the \emph{min-cut condition}.
A set $\D \subset \mathcal{R}^d$ is a feasible region if and only if it satisfies the min-cut condition.

Larger feasible region means more flexibility in finding the suitable repair bandwidth. Therefore, we only need to consider the maximal feasible regions. The following theorem shows that a maximal feasible region can be described by a $k$-tuple $(x_1, x_2, \cdots, x_k)$.
\begin{theorem}
\label{optimalform}
A maximal feasible region $\D$ can be written in the following form:
\begin{eqnarray*}
\D = \{ \pmb{\beta} \, | \sigma_j(\pmb{\beta})\geq x_j, & j=1,\dots,k\},
\end{eqnarray*}
where $0\leq x_1 \leq \cdots \leq x_k \leq \alpha$ and $ \sum_{j=1}^{k}x_j \geq M $.
\end{theorem}
\begin{IEEEproof}
Let $\D'$ be a feasible region satisfying the min-cut condition in (\ref{mc}). As the sequence $\min_{\Beta \in D'} \sigma_j(\Beta)$ is non-decreasing for $j=1,\ldots,k$, we let $i$ be the largest integer such that $\min_{\Beta \in \D'}\sigma_{k-i+1}(\Beta) \geq \alpha$. Thus the min-cut of an information flow graph under $\D'$ is
\[
\sum_{j=1}^{k-i} \min_{\pmb{\beta}\in\D'} \sigma_j(\pmb{\beta})+ i\alpha \geq M.
\]

If $\D'$ cannot be rewritten in the required form, we construct another feasible region $\D''\supset\D'$ as
\begin{eqnarray*}
\D'' = \{\Beta | \sigma_j(\Beta)\geq x_j, j=1,\cdots,k \}
\end{eqnarray*}
\textrm{where}
\begin{eqnarray*}
x_j = \left\{
\begin{array}{ll}
\displaystyle{\min_{\Beta \in \D'} \sigma_j(\Beta),} & j=1,\dots,k-i, \\
 \alpha\textcolor{blue}{,} & j=k-i+1,\dots,k.
\end{array}\right.
\end{eqnarray*}

Since $\D'$ satisfies the min-cut condition, we can deduce that $0\leq x_1\leq \cdots \leq x_k \leq \alpha$ and $\sum_{j=1}^{k}x_j \geq M$. Thus, $\D''$ is of the required form. To complete the proof, it is sufficient to show that $\D''$ is maximal and satisfies the min-cut condition.

From the construction, we have $\D' \subset \D''$, since for any $\pmb{\beta} \in \D'$ it must hold that $\sigma_j(\Beta)\geq x_j$. As $\D'$ is a maximal feasible region, we conclude $\D''$ is maximal. We can verify that $\D''$ also satisfies the min-cut condition:
\begin{eqnarray*}
MC(\D'',\alpha)&=&\sum_{j=1}^k\min\left\{\min_{\pmb{\beta} \in \D''}\sigma_j(\pmb{\beta}), \alpha\right\}\\
&= & \sum_{j=1}^{k-i} \min_{\pmb{\beta}\in\D''} \sigma_j(\pmb{\beta})+ i\alpha \\
&\geq & \sum_{j=1}^{k-i}x_j + i\alpha\\
&=& \sum_{j=1}^{k-i} \min_{\pmb{\beta}\in \D'}\sigma_j(\pmb{\beta})+ i\alpha \quad \geq M
\end{eqnarray*}
\end{IEEEproof}

\subsection{The optimal feasible region for $\alpha = M/k$}
\label{sec: analysis}
For the MSR case where the minimum storage $\alpha = M/k$, we find that there exists one maximum feasible region, which minimizes the regeneration time for any link capacity settings.

\begin{theorem}
For the case of $\alpha = M/k$, any feasible region is a subset of the maximum feasible region
\[\D^* = \{\Beta|\sigma_1(\Beta)\geq M/k\}\]
\end{theorem}
\begin{IEEEproof}
First, it can be seen that for any set $\D$ 
\[
\min_{\Beta \in\D}\sigma_1(\Beta) \leq  \min_{\Beta \in\D}\sigma_2(\Beta) \leq \ldots \leq \min_{\Beta \in\D}\sigma_k(\Beta)
\]
because for any $j=1,2,\cdots,k-1$ 
\[\min_{\Beta\in\D}\sigma_{j+1}(\Beta)  =   \sigma_{j+1}(\Beta')\\
 \geq  \sigma_{j}(\Beta') \label{inequ}\\
 \geq  \min_{\Beta\in\D}\sigma_j(\Beta),
 \]
where $\Beta'$ is the repair bandwidth that minimizes $\sigma_{j+1}(\pmb{\beta}')$. Thus, we have
\[
MC(\D^*, \alpha)  
 \geq  k \min_{\Beta \in\D^*}\sigma_1(\Beta) \geq M
\] 
which means that $\D^*$ is indeed a feasible region. 

On the other hand, for any feasible region $\D$ and any $\Beta \in \D$, if $\sigma_1(\Beta) < M/k$, then
\begin{eqnarray*}
MC(\D, \alpha) < M/k + (k-1)\alpha = M
\end{eqnarray*}
which contradicts the min-cut condition in (\ref{mc}). Thus, we must have $\sigma_1(\Beta)\geq M/k$ and hence $\Beta \in \D^*$.
\end{IEEEproof}

As a conclusion, when using the MSR codes, we can set the feasible region to be $\D^*$ and in each round of repair, we determine the repair bandwidth $\Beta$ by solving the optimization problem (\ref{optimization1}), which is reduced to:
\begin{eqnarray}
\displaystyle \min_{\pmb{\beta}\in \D^*} & &  \max\limits_{i=1,\cdots,d}\frac{\beta_i}{c_i}\label{msrConst1}\\
\textrm{subject to:}  & & \sigma_1(\pmb{\beta})\geq M/k \nonumber
\end{eqnarray}
where $c_i$ is the link capacity from the $i$-th provider $v_i$ to the newcomer $v_0$.

Without loss of generality, we may assume $c_1 \leq c_2 \leq \cdots \leq c_d$. Then the optimal solution $\Beta^*$ to problem (\ref{msrConst1}) can be solved explicitly as follows
\begin{displaymath}
\beta^*_j=\left\{
\begin{array}{ll}
\displaystyle\frac{c_jM}{k\sum_{i=1}^{d-k+1}c_i} & \textrm{if $ 1 \leq j \leq d-k+1 $} \\
\rule{0pt}{18pt}\displaystyle\frac{c_{d-k+1}M}{k\sum_{i=1}^{d-k+1}c_i} & \textrm{if $ d-k+1 < j \leq d  $}
\end{array} \right.
\end{displaymath}

\subsection{The feasible region for $\alpha > M/k$}
\label{sec: algorithm}
For the case of $\alpha>M/k$, we find out that there does not exist a maximum feasible region for any $\alpha > M/k$ and $ k\geq 3$ (Please refer to appendix for a detailed proof). In other words, the optimal feasible region depends on the link capacities. While regenerating data for a new node, we still do not know the link capacities of helper nodes that will join when future failures occur, so we cannot determine the conditions for the optimal solution. Here we use an example to explain this statement.

{\em Example 1:} Set $n=5$, $k=3$, $d=4$, $M=12$, and $\alpha=6$, and consider the following two feasible regions:
\begin{eqnarray*}
\D_1 & = & \{\pmb{\beta}| \sigma_1(\pmb{\beta})\geq 1, \sigma_2(\pmb{\beta})\geq 5, \sigma_3(\pmb{\beta}) \geq 6 \},\\
\D_2 & = & \{\pmb{\beta}| \sigma_1(\pmb{\beta})\geq 2, \sigma_2(\pmb{\beta})\geq 4, \sigma_3(\pmb{\beta}) \geq 6 \}.
\end{eqnarray*}

We find two different repair bandwidths, $\pmb{\beta}_1=(0,1,4,4) \in \D_1 \backslash \D_2$ and $\pmb{\beta}_2=(0,2,2,2) \in \D_2 \backslash \D_1$. If the corresponding link capacities $c_i$ are given as $(1,1,4,4)$, the regeneration times for $\Beta_1$ and $\Beta_2$ are 1 second and 2 seconds, respectively. Under this capacity setting, $\D_1$ is a better solution. However, with another setting of link capacities $(1,2,2,2)$, the regeneration times are then changed to 2 seconds and 1 second, respectively. Feasible region $\D_2$ outperforms $\D_1$ in this setting. According to our previous analysis, $\D_1, \D_2$ are both maximal feasible regions and there does not exist a feasible region including both of them. Thus it is unable to minimize the regeneration times for both the above capacity settings simultaneously.

From this example, we can conclude that, in order to minimize the regeneration time, knowledge on the provider-to-newcomer link capacities for the next rounds of repairs is necessary, which is impractical in real-world distributed storage systems. Therefore, we propose a heuristic feasible region instead:
\[
\D^* = \{\Beta \ | \ \sigma_j(\Beta) \geq \min\{(d-k+j)\beta,\alpha\}, j=1,\cdots,k\}
\]
where $\beta$ is the amount of data downloaded from each provider in the conventional regenerating scheme, which can be calculated according to the optimal tradeoff between storage $\alpha$ and the total repair bandwidth $\sum_{i=1}^d \beta_i$ from $\Beta$ \cite{Dimakis2010}. It can be seen that the repair traffic of conventional regenerating scheme $\Beta = (\beta,\cdots,\beta)$ belongs to $\D^*$. Therefore, with this feasible region, the regeneration time will never be worse than the conventional regenerating scheme.

With the feasible region $\D^*$, we can determine the amount of data to be downloaded from each provider in each round of repair by solving the linear programming problem (\ref{optimization1}).
In case that the solution $\Beta=(\beta_1, \cdots, \beta_d)$  takes on fractional values, its components can be rounded up to their nearest integers. Note that we may choose large $M$ to make the rounding error negligible.

\section{Tree-structured Regeneration with Constant Repair Traffic}
\label{sec:tree}
Li {\em et al.} \cite{tree} first proposed a tree-structured regeneration scheme which transmits the regeneration traffic along a carefully selected tree spanning all the providers. However, as shown in the appendix, this method cannot maintain the MDS property. In this section, we reconsider the problem of optimizing the regeneration time for tree-structured regeneration and maintaining the MDS property featured by entire distributed storage systems.

A tree-structured regeneration solution has two parts: the regeneration tree $T\subset E$ and the number of blocks $f(u,v)$ transmitted on each link of the tree. In the following subsections, we first study the minimum $f(u,v)$ in a given regeneration tree to preserve the MDS property. Thereafter, we show that the problem of building an optimal regeneration tree is NP-hard. Finally, we conclude this section with a heuristic algorithm for constructing a regeneration tree.



\subsection{The minimum $f(u,v)$ for a given regeneration tree}
As the conventional information flow graph is originally constructed for the star topology, we first generalize it to the tree topology for regeneration.

\begin{figure}[!htbp]
\begin{center}
  \subfigure[regeneration tree]{\includegraphics[width=0.21\textwidth]{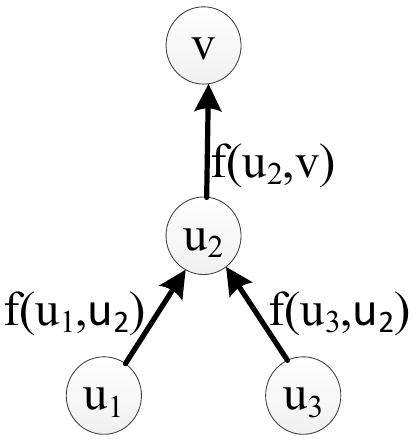}\label{flow_a}}$\quad$
  \subfigure[corresponding part of information flow graph]{\includegraphics[width=0.4\textwidth]{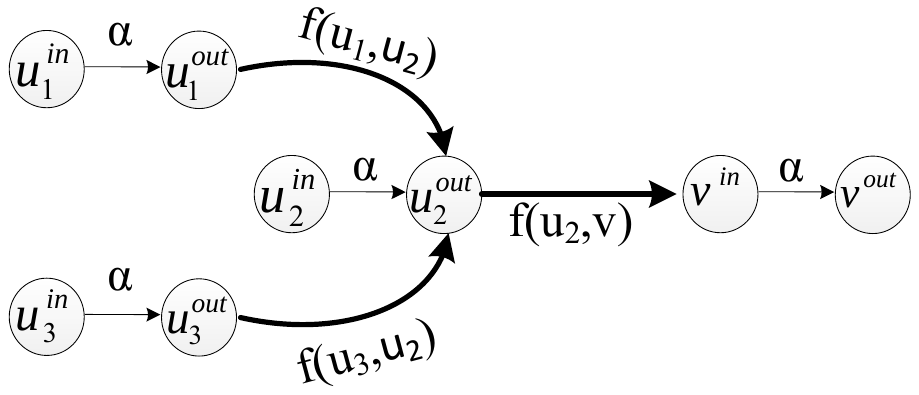}\label{flow_b}}
  \caption{An example of the part of the generalized information flow graph for a regeneration tree consisting of three providers  $u_1, u_2, u_3$ (i.e., $d=3$) and a newcomer $v$.}\label{informationflowgraph}
\end{center}
\end{figure}

To construct the information flow graph for tree-structured regeneration, instead of simply adding $d$ links from the providers' out-nodes to the newcomer's in-node, we rather add links according to the regeneration tree $T$ and the number of blocks $f(u,v)$ transmitted on each link of the tree.
During the regeneration, if provider $u$ transmits $f(u,w)$ blocks to provider $w$, we add a link from $u^{out}$ to $w^{out}$ with capacity $f(u,w)$, whereas if provider $u$ transmits $f(u,v)$ blocks directly to the newcomer $v$, we add a link from $u^{out}$ to $v^{in}$ with capacity $f(u,v)$. For example, Fig.~\ref{flow_a} shows a regeneration tree consisting of three providers $u_1, u_2, u_3$ and a newcomer $v$, and Fig.~\ref{flow_b} presents the corresponding part of information flow graph.

With the generalized information flow graph, we are able to determine the minimum flow $f(u,v)$ on each link that ensures the MDS property.

\begin{theorem}
For a given regeneration tree $T$ rooted at the newcomer, in order to preserve the MDS property, the minimum number of blocks transmitted on each link $(u,v)\in T$ is
\[
\min\{m_u\beta, \alpha\}
\]
where $m_u$ is the number of nodes in the subtree rooted at $u$, and $\beta$ is the number of blocks transmitted in the conventional regenerating scheme, which satisfies $\sum_{i=1}^k \min\{(d-i+1)\beta, \alpha\} = M$.
\end{theorem}

\begin{IEEEproof}
The key is to compute the min-cut of the generalized information flow graph. Let $[U,\bar{U}]$ denote a min-cut that separates the data collector DC from the source, where $U$ is the set of nodes containing DC. Then $U$ must contain at least $k$ ``out'' nodes. Label the $k$ corresponding storage nodes as $u_1, u_2, \cdots, u_k$ in a topological order, such that $u_i$ is a provider in regenerating $u_j$ only if $i<j$. Fig.~\ref{mincut} demonstrates the min-cut in an
information flow graph for a series of tree-structured regeneration processes for $d=4$ and $k=3$, where $u_i$ is a provider with respect to $u_j$ for $1\leq i<j\leq 3$, and $U$ contains all the in-nodes and out-nodes of $u_1,u_2,u_3$.

We divide the cut links into $k$ sets as follows: if $u_i^{in} \notin U$, then the $i$-th set contains only one link $(u_i^{in}, u_i^{out})$; otherwise, the $i$-th set contains all the cut-links introduced in the repair of node $u_i$. According to our scheme, a link $(u, v)$ has flow rate $m_u\beta$ only if the subtree rooted at $u$ has $m_u$ nodes. For repairing $u_i$, as there are at most $i-1$ providers in $U$, there are at least $d-i+1$ providers in $\bar{U}$. Thus the total flow rate of the $i$-th set is no less than $\min \{(d-i+1)\beta, \alpha\}$.
\begin{figure}[!htbp]
\begin{center}
  \includegraphics[width=0.45\textwidth]{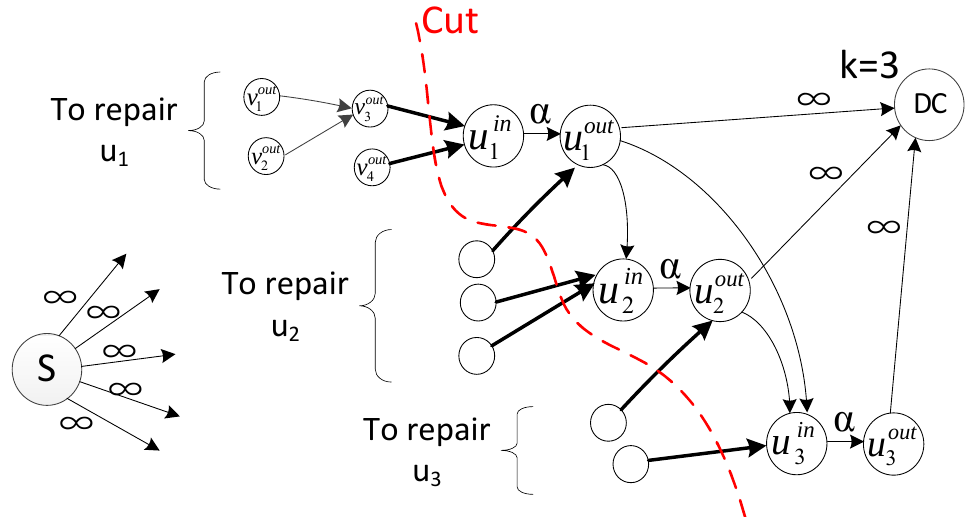}\\
  \caption{An example of min-cut in an information flow graph in the tree-structured regeneration, where $d=4,$ and $k=3$. The dashed line shows the min-cut and the bold links are the cut-links.}\label{mincut}
\end{center}
\end{figure}

Sum up the volumes of the cut links. The total volume of the cut $[U,\bar{U}]$ must be no less than $\sum_{i=1}^k \min\{(d-i+1)\beta, \alpha\} = M$. Thus, if $f(u,v) \geq \min\{m_u\beta, \alpha\}$,  DC can construct the file by assessing any $k$ storage nodes.

To show that we cannot further reduce $f(u,v)$, we need to show that the min-cut with the minimum volume $M$ is achievable. According to the information flow graph constructed in the proof, we can know that such a min-cut is achievable if we require each provider to provide $\beta$ blocks, which means that the solution $\min\{m_u\beta, \alpha\}$ is optimal for a given regeneration tree.
\end{IEEEproof}


\subsection{Construction of the optimal regeneration tree}
\label{sec: ORTproblem}

With the knowledge of how much information to be transmitted on each link, the remaining task is to build an optimal regeneration tree $T$ to minimize the regeneration time. However, we find that the optimal regeneration tree (ORT)  problem is NP-hard.
To demonstrate this, we start with the formal definition of the ORT problem.
\begin{definition}
For a given overlay network $G(V,E)$ with link capacities $c(u,v), (u,v)\in E$, the optimal regeneration tree problem is to find a spanning tree $T$ such that the regeneration time $\max\{\frac{f(u,v)}{c(u,v)} | (u,v) \in T\}$ is minimized, where  $f(u,v)=\min\{m_u\beta,\alpha\}$ and $m_u$ is the number of nodes in the subtree rooted at  $u\in V$.
\end{definition}

To study the complexity of the ORT problem, it is equivalent to restating this optimization problem as a decision problem, which aims to determine whether the regeneration time of an optimal regeneration tree is no more than $1$. The following theorem shows that this problem is NP-hard.
\begin{theorem}
The ORT problem is NP-hard.
\end{theorem}
\begin{IEEEproof}
We first show that ORT $\in$ NP. Suppose we are given a graph $G=(V,E)$. The certificate we choose is the optimal regeneration tree $T$. The verification algorithm checks, for each edge $(u,v) \in T$, that $\frac{f(u,v)}{c(u,v)}\leq 1$. This verification can be performed in polynomial time.

We now prove that the ORT problem is NP-hard by reduction from the VERTEX-COVER problem, which is known to be NP-complete. In particular,  given an undirected graph $G=(V,E)$ and an integer $k$, the VERTEX-COVER problem asks whether all edges can be ``covered'' by $k$ nodes, where node $u\in V$ can cover edge $e\in E$ only if they are adjacent. For an instance of the VERTEX-COVER problem, we construct a regeneration scenario in an overlay network $G'=(V',E')$, such that the regeneration time is less than 1 if and only if $G$ has a vertex cover of size $k$.

We construct $G'$ in the following way. $G'$ has four layers of nodes.
The first layer has only one node that is the root $t$. The second layer has two nodes, node $a$ and node $b$.  Both $a$ and $b$ are connected to the root. The link capacity of edge $(a,t)$ is $k+|E|+1$ and the link capacity of edge $(b,t)$ is unlimited. The nodes in the third layer correspond to the vertices in graph $G$, all of them are connected to both $a$ and $b$. The link capacity of each edge connected to $a$ is unlimited, whereas the link capacity of each edge connected to $b$ is $1$.
The nodes in the last layer correspond to the edges in graph $G$. Each node in the last layer is connected to two nodes in the third layer by the corresponding edges in graph $G$. The edges, which connect the last layer nodes to the third layer nodes, each have an unlimited capacity. Links that are not mentioned in the construction are supposed to have zero capacity.

From the construction above, graph $G'$ can be constructed from $G$ in polynomial time. Fig.~\ref{npc} shows an example of this reduction for the VERTEX-COVER problem with $k=2$, where Fig.~\ref{a} is an instance of $G$, and Fig.~\ref{b} is the graph $G'$ constructed from $G$.

\begin{figure}[h]
\centering
\subfigure[]{\includegraphics[width=4cm,height=3.5cm]{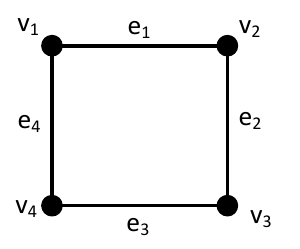}\label{a}}
\subfigure[]{\includegraphics[width=5.2cm,height=5cm]{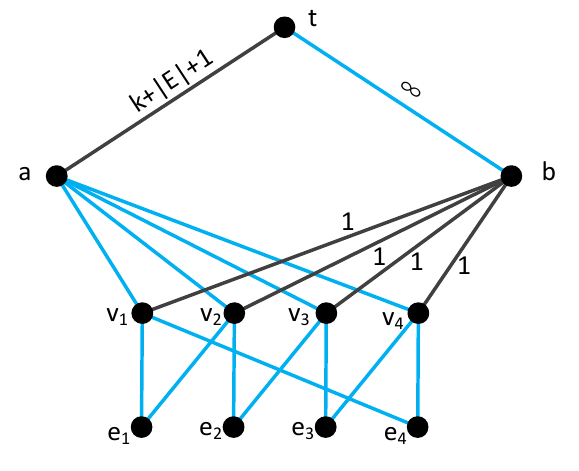}\label{b}}
\caption{Reduction of the VERTEX-COVER problem to the ORT problem for $k=2$: (a) An undirected graph $G=(V,E)$; (b) The graph $G'$ produced by the reduction procedure.}\label{npc}
\end{figure}

We next show that this transformation of $G$ into $G'$ is a reduction. First, suppose that $G$ has a vertex cover set $V'\subseteq V$, where $|V'|=k$. We claim that we can find a tree whose regeneration time is no more than $1$. This tree is constructed according to the vertex cover as follows. For each node $e_i$ of the fourth layer, let its parent be $v_j$ which covers the edge $e_i$ in the vertex cover of $G$. For each node $v_i$ of the third layer, if it belongs to the vertex cover $V'$, let its parent be $a$; otherwise, let its parent be $b$. Finally, let the parent of $a$ and $b$ be the root. It can be verified that the regeneration time is exactly equal to $1$.

Conversely, suppose that $G'$ has a tree whose regeneration time is no more than $1$. Then, we claim that the edges of $G$ can be covered by no more than $k$ nodes. Let $V'\subseteq V$ be the set of nodes  that correspond to children of node $a$ in the regeneration tree. First, $V'$ is a vertex cover of $G$. If it is not the case, some nodes in the fourth layer will have to be connected to the root through $b$, causing some flow entering $b$ larger than $1$ and the regeneration time less than $1$. Second, we show that $|V'| \leq k$. As all the $|E|$ nodes in the fourth layer must transmit their data to the newcomer through $a$, there will be at least $|V'|+|E|+1$ nodes transferring data from node $a$ to root $t$. Because the capacity of link $(a,t)$ is $k+|E|+1$ and the regeneration time is no more than $1$, we conclude that $|V'|\leq k$.
\end{IEEEproof}

\subsection{The heuristic algorithm for constructing the optimal regeneration tree}
In this subsection, we propose a heuristic algorithm to solve the ORT problem since it is NP-hard as mentioned above. The algorithm is inspired by Prim's algorithm \cite{prim1957shortest} for the maximum weighted spanning tree problem. We start from a tree containing only the newcomer as the root and iteratively add the remaining nodes to the tree until it spans all the providers. In each iteration, we try all possible positions for each remaining provider and choose the best position to add the corresponding provider into the regeneration tree. The details are shown in Algorithm~\ref{heuristic}, where $v_0$ represents the root for the newcomer, and $(v',u')$ records the best positions.

\begin{algorithm}
\caption{Find a regeneration tree $T$ for a given network $G(V,E)$.}
\label{heuristic}
\begin{algorithmic}[1]
\STATE Input: Network topology $G(V,E)$, link capacities $c(u,v)$, storage amount $\alpha$, repair traffic $\beta$
\STATE Output: Regeneration tree $T$
\STATE $T \leftarrow v_0$
\STATE $A \leftarrow V-\{v_0\}$
\WHILE {$A \neq \varnothing$}
\FORALL {$v \in A$}
\FORALL {$u \in T$}
\STATE Compute the regeneration time $\max\{\frac{f(u,v)}{c(u,v)}\ | $ $\ (u,v)\in{T\cup \{(v,u)\}}\}$ for tree $T\cup \{(v,u)\}$
\STATE If the regeneration time is better than previous choices, update $(v',u') \leftarrow (v,u)$
\ENDFOR
\ENDFOR
\STATE $T \leftarrow T \cup\{(v',u')\}$
\STATE $A \leftarrow A-\{v\}$
\ENDWHILE
\end{algorithmic}
\end{algorithm}

The most time-consuming step is to test all possible positions for each provider, which has no more than $|V|^2$ choices. Each test takes a linear time of order $O(|V|)$ to compute the regeneration time. Thus, the algorithm runs in polynomial time of order $O(|V|^3)$.

%

\section{Tree-structured Regeneration with Flexible End-to-End traffic}
\label{sec: flexible}
In Sections~\ref{sec:fr} and \ref{sec:tree}, we have discussed two independent approaches to reduce the regeneration time: 1) allowing non-uniform end-to-end repair traffic; 2) allowing tree-structured regeneration topology. In this section, we propose a Flexible Tree-structured Regeneration (FTR) scheme, which combines the advantages of the two approaches to further reduce the regeneration time.

We present the logic flow of the proposed FTR scheme in three steps. First, we analyze the restrictions on the amount of repair data generated by each provider to maintain the MDS property. Second, for a given regeneration tree, we calculate the optimal regeneration time based on the analysis in the first step. Finally, as we are able to determine which one of two trees results in a faster regeneration, we obtain a heuristic algorithm based on local searching.


\subsection{A sufficient condition for the MDS property}

Roughly stated, to maintain the MDS property under a flexible traffic strategy, if the providers connected to low-capacity links generate less coded
blocks, then the providers connected to high-capacity links will have to generate more
coded blocks. Let $\beta_i$ denote the amount of
repair traffic, i.e., the number of coded blocks generated by the
$i$-th provider. Our first task is to analyze the explicit
restrictions on $\beta_i$ that ensures the MDS property.

As analyzed in Section~\ref{sec:tree}, during the
tree-structured regeneration, an intermediate node needs to
re-encode the received blocks from its children only when the number
of received blocks plus the number of blocks generated by itself is
larger than $\alpha$. In this case, the intermediate node transmits
only $\alpha$ coded blocks. Thus, for a regeneration tree $T$, the
number of coded blocks transmitted on each link $(u,v)\in{T}$ will
be
\[
f(u,v)=\min\left\{\sum_{v_i \in S(u)}\beta_i, \alpha\right\}
\]
where $u$ denotes an intermediate node, $v$ denotes
the parent of $u$ in the tree, $S(u)$ denotes the set of nodes in the subtree rooted at $u$, and $\beta_i$ indicates
the number of coded blocks generated at $v_i$ and transmitted to $u$. The following theorem provides a sufficient condition for the MDS property.

\begin{theorem}
The MDS property is
maintained if in each regeneration, we choose
$\beta_i, i=1,2,\cdots,d$, that satisfy
\[
\sum_{l=1}^{d-k+j} \beta_{i_l} \geq \min\{(d-k+j)\beta, \alpha\} \quad \forall j=1,2,\cdots, k
\]
where $\beta$ is again the number of coded blocks
generated by a provider in the conventional regenerating scheme, and
$(i_1,i_2,\cdots,i_d)$ is a
permutation of $(1,2,\cdots,d)$
such that $\beta_{i_1} \leq \beta_{i_2} \leq \cdots \leq
\beta_{i_d}$.\label{thm:ftrlink}
\end{theorem}

\begin{IEEEproof}
We need to prove that any min-cut $[U,\bar{U}]$ (DC $\in U$) has volume of at least $M$.

As the link capacity from a storage node to the
data collector DC is set to be infinity in the
information flow graph, we only need to consider the case that $U$
contains at least $k$ storage nodes. Let $v_1, v_2, \cdots, v_k$ be
the first $k$ storage nodes of $U$ in the topological order.
For $j=1,2,\cdots,k$, if the in-node of $v_j$ is in
the cut, all the links connected to $v_j$ will also be included
in the cut. However, if only the out-node of $v_j$ is in the cut, the link
from the in-node to the out-node of $v_j$ that has capacity
$\alpha$ will be included. 

Following the way we determine the flow
on the regeneration tree, for the $j$-th storage
node $v_j$, the number of providers not in $U$ will be at least $d-j+1$, and
hence the total capacity of the cut links to $v_j$ will
be at least
\[
\sum_{l=1}^{d-k+j} \beta_{i_l} \geq \min\{(d-k+j)\beta, \alpha\}
\]
Therefore, the volume of the cut $[U,\bar{U}]$ will be no less than
$M$, which ensures the MDS property.
\end{IEEEproof}

\subsection{Determination of the optimal regeneration time for a given tree}

To support the non-uniform end-to-end repair
traffic for a tree-structured regeneration, we introduce a
parameter $c_x$ to
denote the end-to-end capacity from provider $x$ to the newcomer $v_0$, where $x$ may not be directly connected to
$v_0$. Let $t$ denote the regeneration time. For a
given regeneration tree $T$, the maximum amount of
data transmitted in $t$ seconds through link $(u,v)$ is $\sum_{x\in S(u)}tc_x$ if we do not
perform encoding at the intermediate node $u$. However, Theorem
\ref{thm:ftrlink} shows that if $\sum_{x\in S(u)}tc_x > \alpha$, we
may encode the repair data at the intermediate node $u$ and transmit
only $\alpha$ blocks to node $v$. Therefore, we obtain the constraints on link capacities as
follows
\[
t c(u,v) \geq \min\left\{\alpha, \sum_{x\in S(u)}tc_x\right\}, \ \ \  \forall (u,v)\in T
\]
Combining the constraints on the link
capacities and the MDS property, we can write the optimal regeneration
time as the following optimization problem:
\begin{equation}
\min \  t \label{eqn:5}
\end{equation}
subject to:
\begin{eqnarray}
&&t \displaystyle c(u,v) \geq \min\left\{\alpha,\sum_{x\in S(u)}tc_x\right\},~\forall (u,v)\in T \\
&& t \sigma_1(\pmb{c}) \geq (d-k+1)\beta \label{eqn:7}
\end{eqnarray}
where $\pmb{c}=\{c_u\  |\  u\in V-v_0\}$, and
$\sigma_1(\pmb{c})$ is, just like in (\ref{eq: model: min-cut}), defined as the sum of the $d-k+1$ smallest components of $\pmb{c}$.

\subsection{The heuristic algorithm}

Upon examining the linear programming (LP) problem in (\ref{eqn:5}), we find that the optimal
$t$ is achieved by taking equality in the constraint (\ref{eqn:7}) as
\begin{equation}
t=\frac{(d-k+1)\beta}{\sigma_1(\pmb{c})}\label{eqn:min:t}
\end{equation}

Substituting (\ref{eqn:min:t}) into the problem (\ref{eqn:5}), we can convert the objective from minimizing the
regeneration time to maximizing $\sigma_1(\pmb{c})$ :
\begin{equation}
\max \sigma_1(\pmb{c})\label{eqn:max}\\
\end{equation}
subject to:
\begin{equation}
c(u,v)\geq\min\left\{\frac{\alpha\sigma_1(\pmb{c})}{(d-k+1)\beta},\sum_{x\in{}S(u)}c_x\right\},~\forall~(u,v)\in{T}\label{eqn:9}
\end{equation}

As there are an exponential number of different
regeneration trees, we cannot enumerate all of them to find the optimal tree.
Instead, we further study the structure of the LP problem
(\ref{eqn:max}) to perform a local search.


Note that the value $\frac{\alpha\sigma_1(\pmb{c})}{(d-k+1)\beta}$ is independent of links $(u,v)\in T$,
 and it is in fact a threshold on
$\sum_{x\in{}S(u)}c_x$.
 We may decompose the constraint
(\ref{eqn:9}) into two parts by enumerating the number of links with
capacity no less than the threshold. Denote this number by $i$. Then the feasible region given by (\ref{eqn:9}) can be divided
into $d+1$ parts, with $\pmb{c}$ in the $(i+1)$-th part ($0\leq
i\leq d$) satisfying:

1) there are exactly $i$ links, of which each has a capacity no less than the threshold;

2) for each of the rest $d-i$ links,
its capacity $c(u,v) \geq \sum_{x\in S(u)}c_x$.

For each $i$, we run Algorithm \ref{elasticAlgorithm} to find a
candidate regeneration tree with $i$ links having capacity no less
than the threshold $\frac{\alpha\sigma_1(\pmb{c})}{(d-k+1)\beta}$
and finally pick up the best candidate as our regeneration
tree.

\begin{algorithm}
\caption{A heuristic algorithm for Flexible Tree-structured Regeneration}
\label{elasticAlgorithm}
\begin{algorithmic}[1]
\STATE Input: $V, i, k$, assuming $v_0\in V$ is the newcomer.
\STATE Output: A regeneration tree $T$ that spans $V$, and a capacity allocation $\pmb{c} = (c_u, u\in V-v_0)$ to maximize the sum of smallest $m$ elements in $\pmb{c}$ under constraints P1 and P2
\STATE Initialize: $T\leftarrow \emptyset, V'\leftarrow \{v_0\}$
\FOR{$i'=1$ to $i$}
\STATE Among the links in the cut $[V', V-V']$, find the link $(u,v)$ with the largest capacity (assuming $v\in V', u\notin V'$).
\STATE $T \leftarrow T \cup \{(u,v)\}$
\STATE $V' \leftarrow V' \cup \{u\}$
\ENDFOR
\STATE Let $d'=d-i = |V-V'|$, $m=d-k+1$
\FOR{each $u\in V-V'$}
\STATE Let $v'$ be the node maximizing $c(u,v)$ among nodes $v\in V'$
\STATE $c_u \leftarrow c(u,v')$
\STATE $T \leftarrow T \cup \{(u,v')\}$
\ENDFOR
\REPEAT
\STATE Sort $\pmb{c}$ in ascending order $c_{u_1}\leq c_{u_2} \leq \cdots c_{u_{d'}}$;
\STATE For each $j>m$, set $c_{u_j}$ to $c_{u_m}$
\FOR{each $u \in V-V'$}
\STATE Let $(u,v)\in T$ be the link leaving $u$
\FOR{each $v'\neq v$}
\STATE $T'' \leftarrow T - (u,v) + (u,v')$
\IF{$\pmb{c}$ is feasible in $T''$ and $\exists 1\leq i \leq m$, $c_{u_i}$ can be increased}
\STATE increase $c_{u_i}$ to maximum possible under constraints P1 and P2
\STATE $T \leftarrow T''$
\STATE break;
\ENDIF
\ENDFOR
\ENDFOR
\UNTIL{$T$ is not updated in the last loop}
\end{algorithmic}
\end{algorithm}

Algorithm \ref{elasticAlgorithm} finds the candidate tree in two
steps. First, through lines 3--8, we find a
connected subtree containing $i$ links so that the smallest link
capacity of $i$ links is maximized. After the first step, $V'$ is
the set of $i$ providers connected to the newcomer by the subtree.
In the second step, we connect the rest $d'=d-i$ providers to the
newcomer through a local search (lines 10--14), where we start from an initial regeneration tree, and
during each iteration of the loop from line 15 to line 29, we check if we can get a better regeneration tree by a pivot operation that cuts off a subtree and connects it to some other node.

Note that for the case $i=0$, lines 10--14 will set the initial regeneration tree as the star topology. Therefore, the regeneration tree returned
by FTR is always no worse than the FR solution.

\section{Evaluation}
\label{sec: evaluation}

In this section, we present simulation results to verify the effectiveness of our proposed schemes: Flexible Regeneration (FR), Tree-structured Regeneration (TR), and Flexible Tree-structured Regeneration (FTR).
Our most concern is the regeneration time, which is measured as the time that the newcomer spends on regenerating the coded blocks. In the evaluation of the regeneration time, we ignore the encoding time on each provider and the decoding time on the newcomer, because the encoding and decoding operations can be performed simultaneously during the transmission of repair data \cite{tree}.


For default settings, we use the same experiment setup as \cite{tree}, where redundant data is produced using an $(n=20,k=5)$-MDS code. The original file size is set to be $M=1$GB. The link capacities between the storage nodes obey the uniform distribution within the range [10Mbps,120Mbps], which is obtained from the measurement of real-world networks \cite{measuring}.

\subsection{Effect of the number of providers $d$}


\begin{figure*}[htbp!]
\centering
\subfigure[Normalized regeneration time.
]{\includegraphics[width=0.44\textwidth]{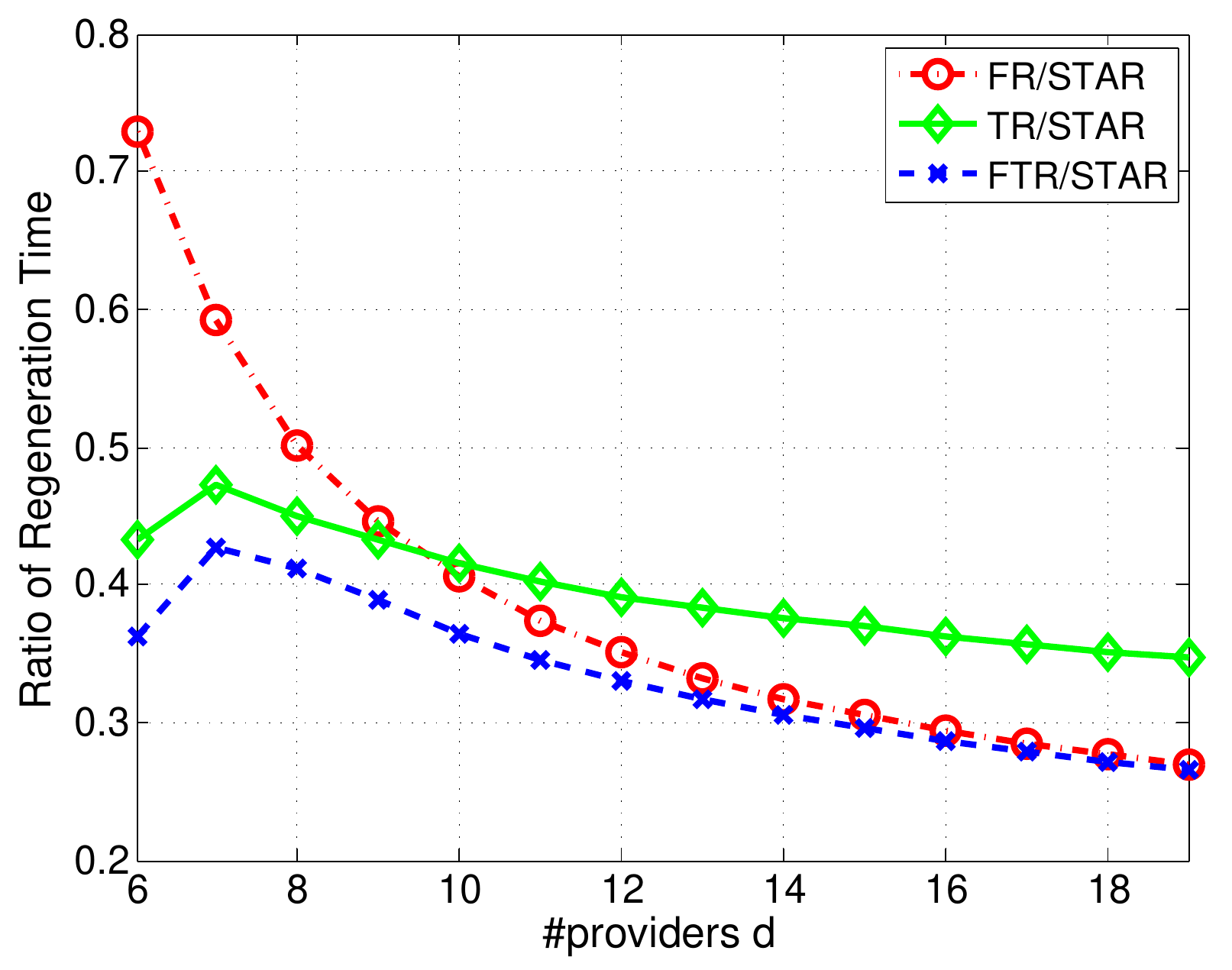}\label{ratio7}}
\subfigure[Normalized bandwidth consumption.
]{\includegraphics[width=0.462\textwidth]{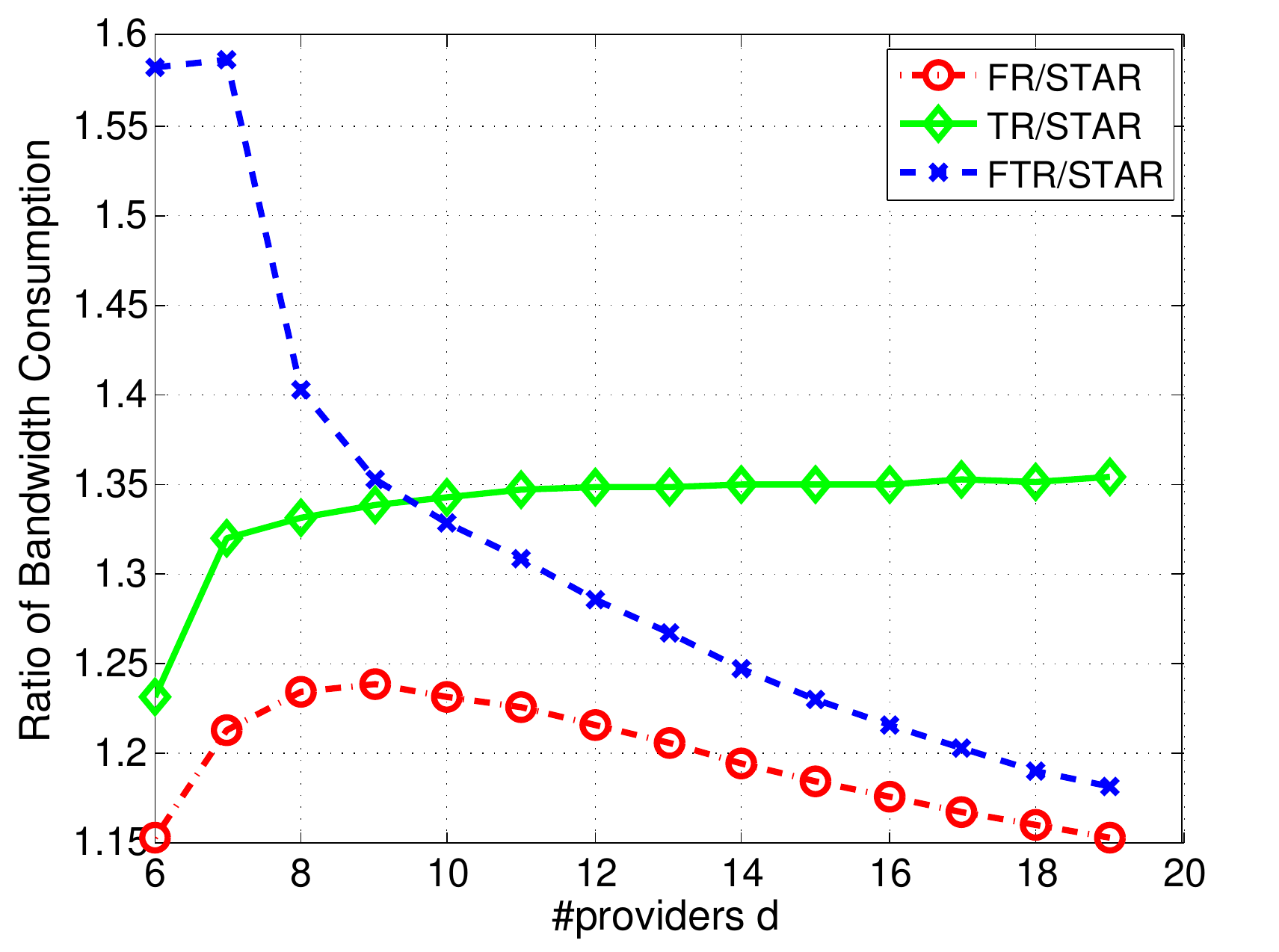}\label{bandwidth7}}
\caption{Effects of $d$ on the performances of the
FR, TR and FTR schemes, in comparison with the STAR scheme based on
uniform repair traffic, for $n=20$, $k=5$, $M=1$GB, and uniform
capacity distribution range [10Mbps,120Mbps].
}\label{performance7}
\end{figure*}

The number of providers $d$ is a key parameter for
regeneration in distributed storage systems.
In the STAR topology, the total repair bandwidth consumed in the regeneration
process decreases as $d$ grows \cite{Dimakis2010}.
In
the case of one node failure, the theoretical optimal value of $d$
is $n-1$ for achieving a minimum repair bandwidth,
although accessing a large number
of providers will introduce extra communication overheads.
On the other hand, all feasible values of $d$ may appear in
practice. In the evaluation, we vary $d$ from $k+1$ to $n-1$, in
order to find out how this factor affects the performance of each
regeneration scheme.

We consider the MSR point, where each node stores
$\alpha=M/k=200$MB.
Fig.~\ref{performance7} presents the simulation
results on performance improvements of the FR, TR, and FTR schemes
with respect to the STAR scheme, where all possible values of $d$
are considered, and the uniform capacity distribution is within the
range [10Mbps,120Mbps]. Note that the regeneration schemes
(STAR) proposed by Dimakis {\em et al.} in \cite{Dimakis2010} is
implemented as a benchmark. For convenience, the
traditional STAR schemes based on uniform repair traffic are simply
referred to as `STAR' below.


In Fig.~\ref{ratio7}, we normalize the regeneration times
of FR, TR and FTR by the regeneration time of STAR to show the
relative improvement. In most cases, our schemes reduce the regeneration time by $50\%
\sim 70\%$ compared with STAR. We note that the
regeneration times of both STAR and the three proposed regeneration schemes all decrease as $d$ grows because of the reduction of total
amount of repair data. Meanwhile, the regeneration times of our
schemes reduce faster than that of STAR. This is because the star
topology has a large chance to include a low capacitated
provider-to-newcomer link when $d$ is large. However, the bottleneck effect can be
alleviated by FR and the tree topology.

An interesting observation is that FR outperforms TR for
large $d$ values, but on the contrary, TR outperforms FR when $d$ is
small. The reason is that, in order to bypass the bottleneck link
with a tree topology, the intermediate provider nodes have to transmit more
data. This effect can only be recovered by raising the minimum capacity for links
connected to a big number of participating providers.

In Fig.~\ref{bandwidth7}, we examine the total repair bandwidth
consumption for the FR, TR, and FTR schemes, where again the repair
bandwidth is normalized by that for STAR. As a tradeoff, our schemes
all sacrifice the repair bandwidth for reducing the regeneration
time. However, it is not surprising that tree-structured
regeneration has higher repair bandwidth consumption than STAR-structured regeneration. For example, in a regeneration tree the amount of repair data is counted
twice if it is transmitted to the newcomer by two hops.


As a conclusion, FTR is always advantageous
than both FR and TR in any case, which is promised by the design of
FTR. When $d$ is large, however, FR has a regeneration time almost
as good as FTR, but enjoys a slightly smaller repair bandwidth.

\subsection{Effect of the bandwidth variance}

\begin{figure}[h]
\centering
  \includegraphics[width=0.48\textwidth]{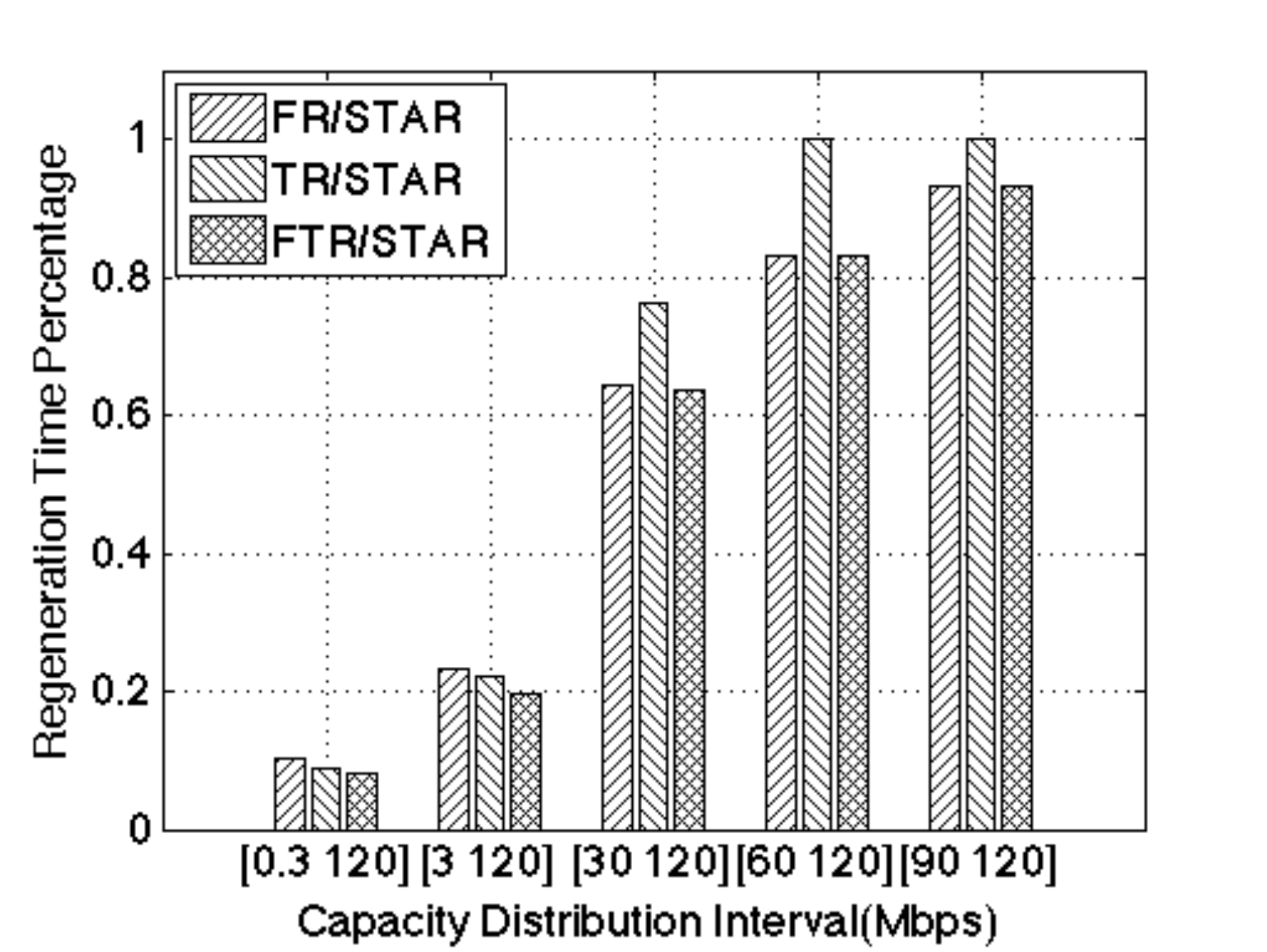}\\
\caption{Effects of network bandwidth on the
regeneration time for the FR, TR, and FTR schemes, where $n=20$,
$k=5$, $d=10$, and $M=1$GB.}\label{d10}
\end{figure}

In order to show the impacts of network bandwidth variance on the regeneration time, we run simulations with 5
different link capacity distributions: $U_1[0.3,120]$Mbps,
$U_2[3,120]$Mbps, $U_3[30,120]$Mbps, $U_4[60,120]$Mbps,
$U_5[90,120]$Mbps. Fig.~\ref{d10} shows the results
with the number of providers $d$ fixed at $10$. The performances of our
schemes are better when the variance of network bandwidth is large.
For uniform distribution $U_1[0.3,120]$Mbps, FR, TR and FTR all achieve a
reduction of about $90\%$ in the regeneration time compared with
the traditional STAR scheme. When the variance of network bandwidth
becomes small, for example at $U_4[60,120]$Mbps and
$U_5[90,120]$Mbps, TR has the same regeneration time as STAR, but
FTR still reduces the regeneration time by $10\% \sim 20\%$.

\subsection{Effect of the storage capacity per node $\alpha$}

Our tests above focus mainly on the MSR point, which
achieves the optimal storage efficiency. However, as shown by
Dimakis {\em et al.} \cite{Dimakis2010}, the repair bandwidth can be
reduced by storing more data on each storage node. To the other
extreme, the MBR point achieves the minimum repair bandwidth.

\begin{figure}
\centering
\subfigure[Normalized regeneration time.]{\includegraphics[width=0.46\textwidth]{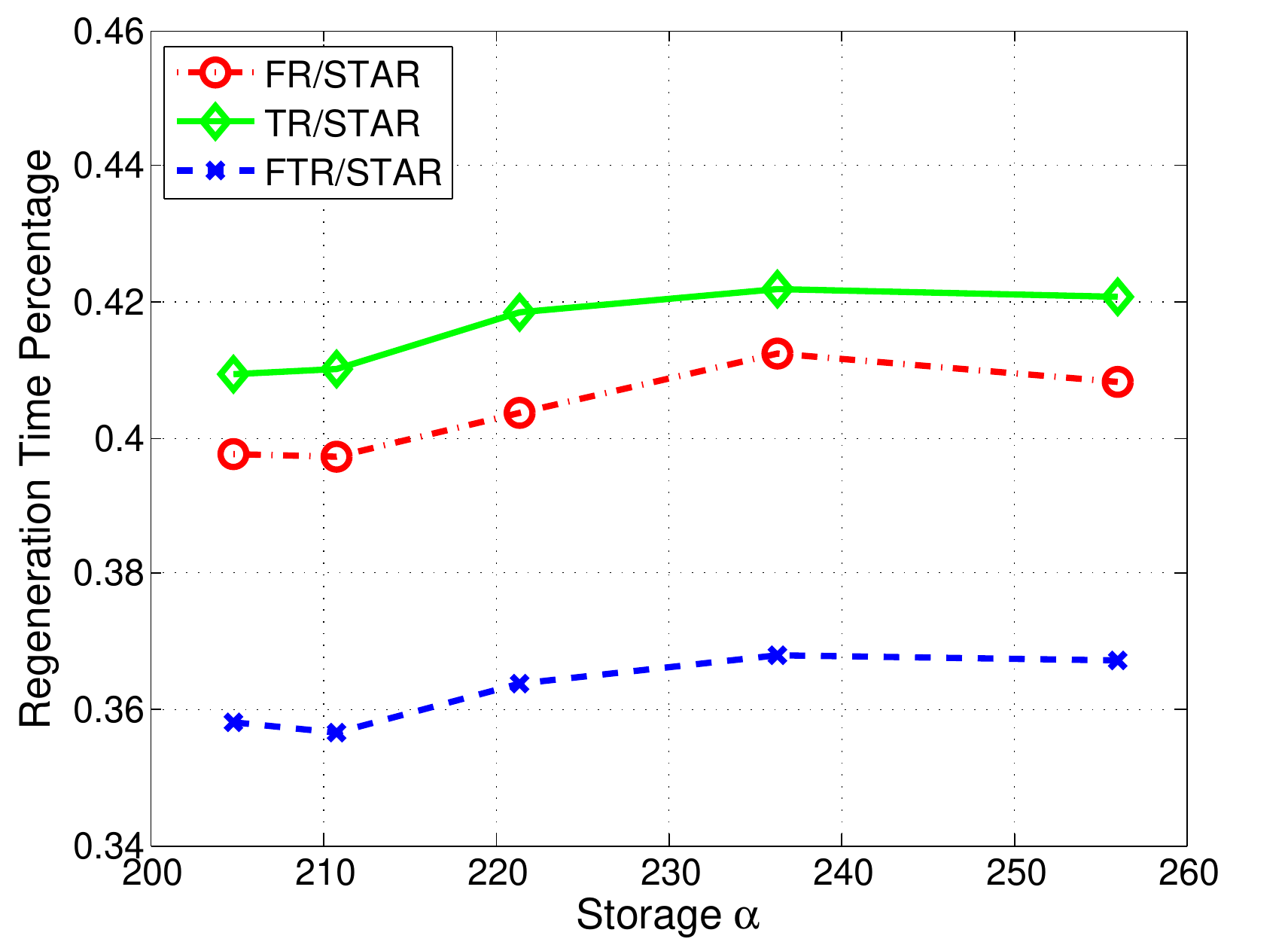}\label{ratio}}
\subfigure[Normalized bandwidth consumption.]{\includegraphics[width=0.46\textwidth]{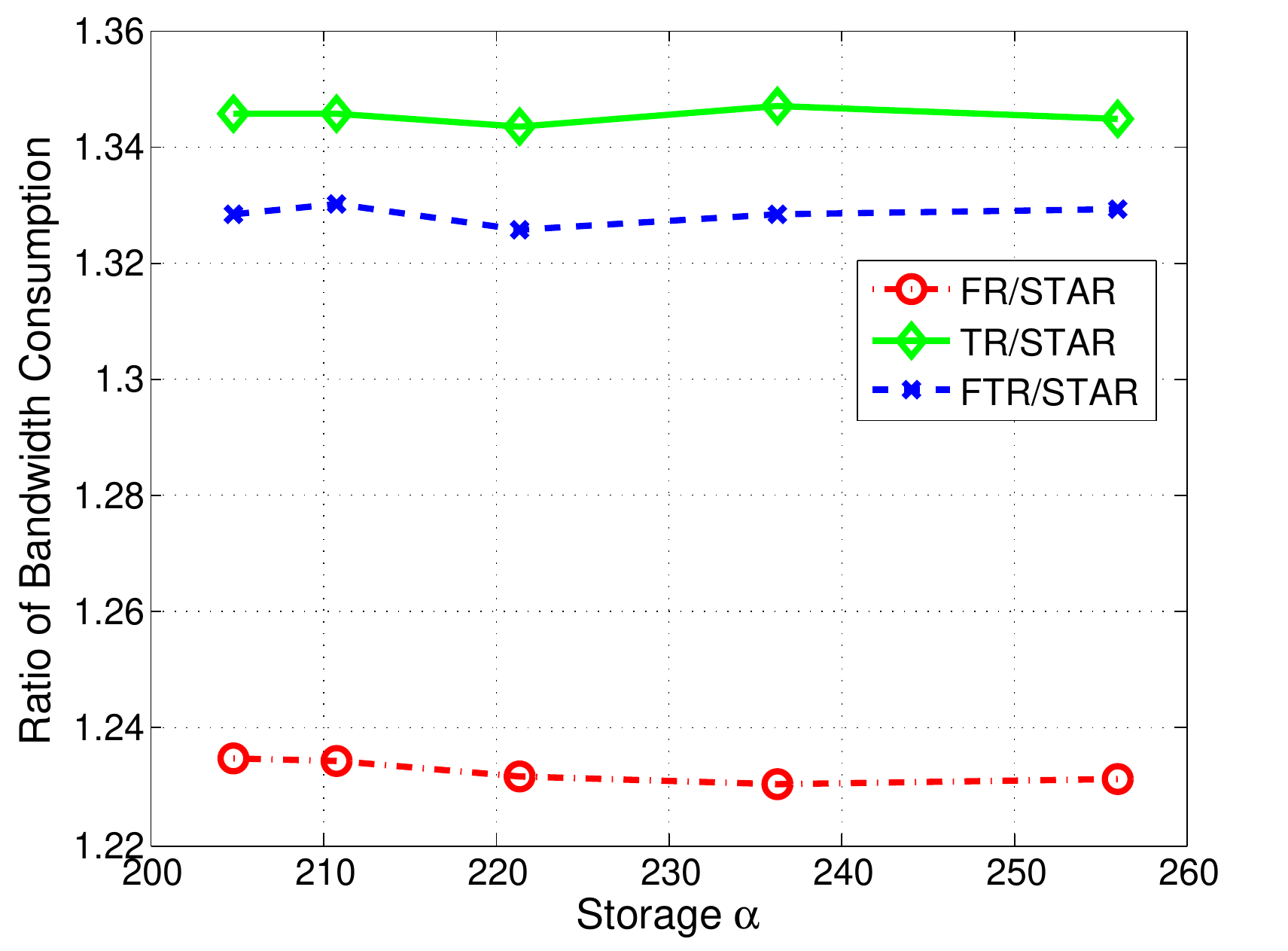}\label{bandwidth}}
  \caption{Effects of storage amount on the regeneration time for the  FR, TR and FTR schemes, where $n=20, k=5, d=10$, $M=1$GB, and $\alpha$ varies from the MSR point to the MBR point.}\label{change_alpha}
\end{figure}

To test the effects of the storage capacity per node
on our schemes, we vary the values of $\alpha$ from the MSR point to the MBR point in the simulations. Fig.~\ref{change_alpha}
shows the results on regeneration times and repair
bandwidths for different $\alpha$ values.
We find that the regeneration time in
each of our scheme does not change much as $\alpha$ varies. This
implies that our previous conclusions for the MSR case also apply to any
non-MSR case with a different storage amount
$\alpha$.


\section{Related Work}
\label{sec: relate}
%

Li {\em et al.} \cite{treeIWQOS} first considered the heterogeneity
of network bandwidth in data regeneration process and proposed a
tree-structured regeneration scheme to reduce the regeneration time.
They also proposed a scheme of building parallel regeneration trees
to   further reduce the regeneration time in the network with
asymmetric links \cite{treeParallel}. However, they only discussed
the case that the regeneration scheme requires $k$ providers, which
means the minimal regeneration traffic is equal to the size of
original file $M$. To further reduce the regeneration time, they
considered the regenerating codes in the tree-structured
regeneration scheme and proposed RCTREE in \cite{tree}. They employ
a minimum-storage regenerating (MSR) codes in RCTREE, which means
that the minimal regeneration traffic is
$\frac{d}{k(d-k+1)}M$ bytes. Therefore, for a regeneration with $d$
providers, each provider sends $\frac{\alpha}{d-k+1}$ blocks to its
parent node. To make sure that the newcomer has enough information
to restore $\alpha$ blocks, it has to receive data directly from at
least $d-k+1$ providers. The details of how to construct an optimal
regeneration tree can be found in Algorithm 1 of
\cite{tree}. Although their algorithm ensures that the degree of newcomer is at least
$d-k+1$, the MDS property still cannot be preserved after data regeneration.

Sun {\em et al.} \cite{treeMultipleDataLosses} considered the
scenario of repairing multiple data losses, and proposed two
algorithms based on tree-structured regeneration to reduce the
regeneration time. However, they assumed the same
amount of data transferred between providers and newcomer for regenerating codes. 
According to our analysis, their regeneration schemes also cannot preserve the MDS property. 

Some researches, such as \cite{heterogeneity,hero}, considered the
heterogeneity of nodes availability and optimized the erasure code
deployment to reduce the data redundancy. Moreover, other
researches, such as
\cite{Akhlaghi2010,Gerami2011a,Downloadcost,communicationcostArmstrong,FlexibleRC_ISIT2010},
jointly considered the repair-cost and heterogeneity of
communication(download) cost on each links. They flexibly determine
the amount of data to minimize the total repair cost, which is
different from the regeneration time.

Regenerating codes suppose that all storage nodes store the same
amount of data and the newcomer obtains the same amount of data from
each provider. However, the communication cost of each provider may
be different. Akhlaghi {\em et al.} \cite{Akhlaghi2010} proposed a
cost-bandwidth trade-off by introducing two classes of storage nodes
with two different communication costs. However, the newcomer may
only contact a determined number of providers in each of the two
classes, and the amount of data downloaded from providers in the
same class remains unchanged. Gerami {\em et al.} \cite{Gerami2011a}
considered the impact of the network topology and proposed the
optimal-cost regenerating codes with variable link costs of
providers with a given network topology. They
assumed that, just like for conventional regenerating codes, newcomer downloads the same
amount of data blocks from each provider.

The generalized repair method with various amount of
information downloaded from each provider was studied by Soroush
Akhlaghi {\em et al.} \cite{Downloadcost}, Craig Armstrong {\em et
al.} \cite{communicationcostArmstrong} and Nihar B. Shah {\em et
al.} \cite{FlexibleRC_ISIT2010}. Armstrong {\em et al.}
\cite{communicationcostArmstrong}  proposed necessary conditions for
the minimum repair bandwidth of the first two repairs at the MSR
point. They conjectured that their result holds for
any number of repairs, which has been proved in this paper. They
also generalized their work to heterogenous storage capacities.
Nihar B. Shah {\em et al.} \cite{FlexibleRC_ISIT2010} proposed a
flexible class of regenerating codes in support of both flexible
reconstruction and flexible regeneration. They accomplished flexible
regeneration with two parameters $\gamma$ and $\beta_{\text{max}}$,
such that the repair bandwidth of each provider can be flexibly
chosen from $[0,\beta_{\text{max}}]$ as long as the total repair
bandwidth is no less than $\gamma$. Their method is
generalized in this paper by introducing the concept of feasible
region, which characterizes the set of feasible repair bandwidth
vectors. We have also compared their work with ours in the
evaluation section above.

\section{Conclusion}
\label{sec: conclusion}

We have reconsidered the problem of how to reduce the regeneration time in
networks with heterogeneous link capacities. We have analyzed the
minimum amount of data to be transmitted on each link of the
regeneration tree, and proved that the problem of building optimal
regeneration tree is NP-complete. Using a proposed heuristic algorithm to
construct a near-optimal regeneration tree, the
regeneration time can further be reduced by
allowing non-uniform end-to-end repair traffic. With
the non-uniform end-to-end repair traffic, we can flexibly determine
the amount of coded data generated by each provider. Simulation
results have shown
that our regeneration schemes are able to maintain the MDS property
and reduce the regeneration time by 10\% $\sim$ 90\%, compared with
traditional star-structured regenerating codes. The proposed Flexible Tree-structured
Regeneration scheme performs even better than RCTREE.

\appendices
\section{Loss of the MDS property in RCTREE}
\label{sec: MDS property}


Let us employ an example to demonstrate that the
RCTREE scheme is unable to maintain the MDS property. Consider the
overlay network shown in Fig.~\ref{example}(a). Assume
that a file of size $M=480$Mb is distributed using a $(n=5,k=2)$ MDS-code to the five storage nodes $v_1, v_2,
\cdots, v_5$, of which each holds $\alpha
=M/k= 240$Mb data. The MDS property requires that the file can be reconstructed by
any two storage nodes. Suppose that $v_5$ fails,
$v_0$ is selected as the newcomer,
and $v_1, \cdots, v_4$ are the $d=4$ providers. In this example,
RCTREE will use the same regeneration tree as shown in
Fig.~\ref{example}(d), where a fixed amount of
$\beta=\frac{M}{k(d-k+1)}=80$Mb is transmitted on
each link for regeneration at $v_0$.

\begin{figure}[h]
  \centering
  \includegraphics[width=0.45\textwidth]{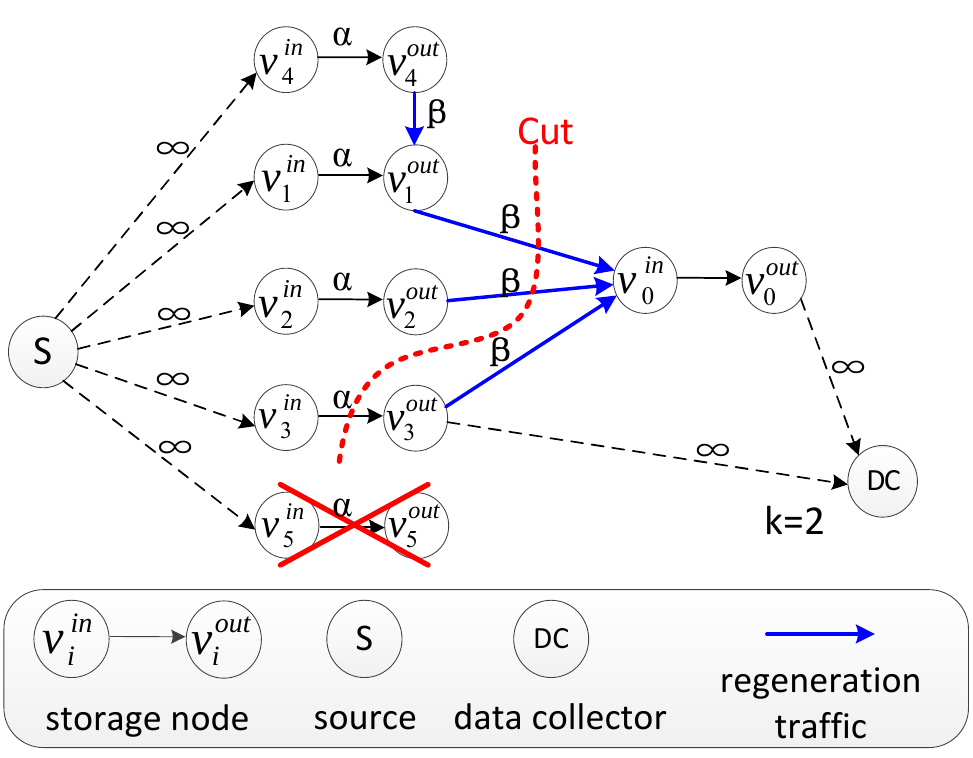}\\
  \caption{An example of RCTREE and the corresponding information flow graph. The parameters are $n=5, d=4, k=2, M=480$Mb, $\alpha=M/k=240$Mb, $\beta$=$\frac{M}{k(d-k+1)}=80$Mb. A min-cut with capacity $2\beta + \alpha=400$Mb is illustrated by the red dashed line, implying that a DC can not reconstruct the original file by connecting with $\{v_3, v_0\}$. }\label{rctree}
\end{figure}

Assume that the data collector connects to $v_0$
and $v_3$ to reconstruct the file. Fig.~\ref{rctree} shows the
information flow graph. As marked in the figure, there is a cut of volume
$2\beta + \alpha = 400$Mb, which is smaller than the file size.
Therefore, the file cannot be reconstructed with storage nodes $v_3,
v_0$.

\begin{figure}[h]
  \centering
  \includegraphics[width=0.45\textwidth]{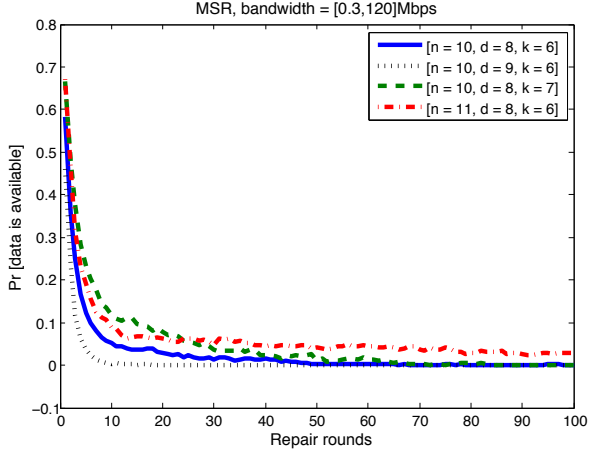}\\
  \caption{Impacts of the number of repair rounds on the probability of successfully reconstructing the original file.}\label{rctreeMDS}
\end{figure}

To find out how frequently the file reconstruction
fails, we have implemented the RCTREE scheme based
on Random Linear Regenerating Codes (RLRC) and run simulations with
practical parameter values.
The finite field $GF(2^{16})$ has been chosen for
RLRC since it is sufficiently large such that the probability
that linearly dependent blocks are regenerated is negligible
\cite{practical}. Fig.~\ref{rctreeMDS} presents the
simulation results for four sets of code parameter settings, showing
the probability of successful file reconstruction as a function of
the number of repair rounds. From this figure, we can see that
 the original file can hardly be reconstructed after 5 repair rounds as the number of original storage
nodes becomes in turn smaller than $k$.

From these
results, we may state that the problem of optimizing
regeneration time with heterogeneous link capacities
should be solved with a satisfaction to
 the MDS property.

\section{The maximal feasible region for the Non-MSR case}

\begin{theorem}
For the non-MSR case of $\alpha > M/k$ and $k\geq
3$, there does not exist a maximum feasible region $\D$.
\end{theorem}

\begin{IEEEproof}
Recall that $\sigma_j(\pmb{\beta})$ is defined as
the sum of the $d-k+j$ smallest components of the repair bandwidth
$\pmb{\beta}=(\beta_1,\beta_2,\cdots,\beta_d$). For a feasible
region $\D$, it always holds that
$\displaystyle\min_{\pmb{\beta}\in\D}\sigma_{j+1}(\pmb{\beta})\geq\min_{\pmb{\beta}\in\D}\sigma_j(\pmb{\beta})$
for $j=1,\cdots,d-1$. Let $i$ denote the number of terms
$\displaystyle{\min_{\pmb{\beta}\in\D}\sigma_j(\pmb{\beta})\geq\alpha}$,
{\em i.e.,}
\[
\min_{\pmb{\beta}\in\D}\sigma_{k-i}(\pmb{\beta}) < \alpha \quad
\land \quad \min_{\pmb{\beta}\in\D}\sigma_{k-i+1}(\pmb{\beta}) \geq
\alpha,
\]
where $i$ must range from $1$ to $k$, since
$\sigma_k(\pmb{\beta})$ must be no less than $\alpha$ for a
successful repair. Therefore, all feasible regions can be partitioned into $k$
groups by the value of $i$, and for $\alpha > M/k$, every group is
non-empty. In order to prove this theorem, it is sufficient to
show that in the $i$-th group, where $1\leq i\leq k-2$ and
$M-i\alpha >0$, there does not exist a maximum region. Note that we call a feasible region {\em maximum}, if it includes all feasible regions. A feasible region $\D$ is {\em maximal}, if adding any vector $\pmb{\beta}\in \mathbb{R}^d \backslash \D$ to $\D$ makes it infeasible.

We prove this by contradiction. If there exists a maximum region $\D_{max}$ in the $i$-th
group, it must contain all the vectors $\pmb{\beta}$ satisfying the
following constraints:
\begin{eqnarray*}
\sum_{j=1}^{k-i} \sigma_j(\pmb{\beta}) & \geq & M-i\alpha \\
\sigma_{k-i}(\pmb{\beta}) & < & \alpha \\
\sigma_{k-i+1}(\pmb{\beta}) & \geq & \alpha \\
\forall j : \beta_j & \leq & \alpha.
\end{eqnarray*}
Then it is sufficient to prove that $\D_{max}$ does not satisfy the min-cut condition.

Pick up a $\pmb{\beta} \in \D_{max}$ and assume that $\beta_1\leq \beta_2 \leq \cdots \leq
\beta_d$ without loss of generality. Under this assumption,
$\sigma_j(\pmb{\beta}) = \beta_1 + \cdots + \beta_{d-k+j}$. If
$\displaystyle\sum_{j=1}^{k-i} \sigma_j(\pmb{\beta})
> M-i\alpha $, we construct $\pmb{\beta}^0$ as follows:

\begin{displaymath}
\beta_j^0=\left\{
\begin{array}{ll}
t\beta_j & \textrm{if $ 1 \leq j \leq d-i $} \\
\alpha & \textrm{if $ d-i+1 \leq j \leq d  $}
\end{array} \right.
\end{displaymath}
where $t=\frac{M-i\alpha}{\sum_{j=1}^{k-i}
\sigma_j(\pmb{\beta})}$. Therefore,
\begin{eqnarray*}
\sum_{j=1}^{k-i} \sigma_j(\pmb{\beta}^0)=t\sum_{j=1}^{k-i} \sigma_j(\pmb{\beta}) & = & M-i\alpha \\
\sigma_{k-i}(\pmb{\beta}^0) < \sigma_{k-i}(\pmb{\beta}) & < & \alpha \\
\sigma_{k-i+1}(\pmb{\beta}^0) & \geq & \alpha \\
\forall j : \beta_j^0 & \leq & \alpha.
\end{eqnarray*}
Thus, $\pmb{\beta}^0\in \D_{max}$. Let $m$ be the minimum integer such that $\sigma_m(\pmb{\beta}^0)>0$. As
$M-i\alpha>0$, we have $m\leq k-i$.
Thus, it is in turn sufficient to prove
\[\sum_{j=1}^{k-i}\min_{\pmb{\beta} \in
\D_{max}}\sigma_j(\pmb{\beta})<
\sum_{j=1}^{k-i}\sigma_j(\pmb{\beta}^0) = M-i\alpha.\]
Because
$\displaystyle\min_{\pmb{\beta}\in\D_{max}}\sigma_j(\pmb{\beta})
\leq \sigma_j(\pmb{\beta}^0)$, we will show that
$\displaystyle\min_{\pmb{\beta}\in\D_{max}}\sigma_m(\pmb{\beta}) <
\sigma_m(\pmb{\beta}^0)$ to complete the proof. Due to the
definition of $m$, we have $\beta_1=\beta_2= \dots
=\beta_{d-k+m-1}=0$, and $\beta_{d-k+m}>0$. Then we construct
$\pmb{\beta}'$ as follows:

\begin{displaymath}
\beta'_j=\left\{
\begin{array}{ll}
\beta_j^0-\epsilon & \textrm{if $ j = d-k+m  $}\\
\beta_j^0+(k-i)\epsilon & \textrm{if $j = d-k+k-i  $}\\
\beta_j^0 & \textrm{otherwise}
\end{array} \right.
\end{displaymath}
where
$0<\epsilon<\min\{\frac{\alpha-\sigma_{k-i}(\pmb{\beta}^0)}{k-i-1},\beta^0_{d-k+m}\}$.

Therefore,
\begin{eqnarray*}
\sum_{j=1}^{k-i} \sigma_j(\pmb{\beta}') & = &
\sum_{j=1}^{m-1}\sigma_j(\pmb{\beta}^0)+\sum_{j=m}^{k-i-1}(\sigma_j(\pmb{\beta}^0)-\epsilon)\nonumber\\
& &+\sigma_{k-i}(\pmb{\beta}^0) \nonumber+(k-i-1)\epsilon \\
 &\geq & \sum_{j=1}^{k-i} \sigma_j(\pmb{\beta}^0)  =  M-i\alpha\\
\sigma_{k-i}(\pmb{\beta}') & = & \sigma_{k-i}(\pmb{\beta}^0)+(k-i-1)\epsilon  < \alpha \\
\sigma_{k-i+1}(\pmb{\beta}') & = & \sigma_{k-i+1}(\pmb{\beta}^0)+(k-i-1)\epsilon\geq  \alpha \\
\forall j : \beta'_j & \leq & \alpha,
\end{eqnarray*}
which means $\pmb{\beta}' \in \D_{max}$ and
$\displaystyle\min_{\beta\in\D_{max}}\sigma_m(\pmb{\beta}) \leq
\sigma_m(\pmb{\beta}')=\sigma_m(\beta^0)-\epsilon$.
\end{IEEEproof}


\bibliographystyle{IEEEtran}
\bibliography{treeref}


\end{document}